\documentclass[12pt]{article}
\usepackage{amssymb}
\usepackage{amsfonts}
\usepackage{amsmath}
\usepackage[dvips]{graphicx}

\begin{document}

\title{Non-local $\mathcal{PT}$-symmetric potentials in the one-dimensional
Dirac equation}
\author{Francesco Cannata$^1$ and Alberto Ventura$^2$ \\
$^1$Istituto Nazionale di Fisica Nucleare, Sezione di Bologna,\\
Via Irnerio 46, I-40126 Bologna, Italy\\
$^2$Ente per le Nuove Tecnologie, l' Energia e l' Ambiente,\\
Via Martiri di Monte Sole 4, I-40129 Bologna,\\
and Istituto Nazionale di Fisica Nucleare, Sezione di Bologna, Italy\\
Francesco.Cannata@bo.infn.it, Alberto.Ventura@bologna.enea.it}
\date{}
\maketitle

\begin{abstract}
The Dirac equation in (1+1) dimensions with a non-local $\mathcal{PT}$%
-symmetric potential of separable type is studied by means of the Green
function method: properties of bound and scattering states are derived in
full detail and numerical results are shown for a potential kernel of
Yamaguchi type, inspired by the treatment of low-energy nucleon-nucleon
interaction.\newline
\end{abstract}


\section{Introduction}

Since the pioneering papers by Bender and coworkers\cite{BB98},\cite{BBM99},
the study of non-Hermitian Hamiltonians invariant under space-time
reflection has developed into a branch of quantum mechanics in its own,
called $\mathcal{PT}$-symmetric quantum mechanics. The large majority of
analyses have been devoted to bound state problems, where the observation
that $\mathcal{PT}$-symmetric Hamiltonians with eigenfunctions that are
eigenstates of $\mathcal{PT}$ \ have real spectra has led to
Hermitian-equivalent formulations, where one can define a linear operator $%
\mathcal{C}$, commuting with Hamiltonian $H$ and with $\mathcal{PT}$ , that
permits constructing time-independent inner products with positive-definite
norms of the form $\int_{-\infty }^{+\infty }\Psi \left( x\right) \mathcal{%
CPT}\Psi \left( x\right) dx$ (see Ref.\cite{Be07} for a review).

$\mathcal{PT}$-symmetric quantum mechanics has a close connection with the
more general quasi-Hermitian quantum mechanics\cite{SGH92},\cite{Mo02a},\cite%
{Mo08}, where $H$ is called quasi-Hermitian if it satisfies the intertwining
relation $H^{\dagger }=\eta _{+}H\eta _{+}^{-1}$, with $\eta _{+}$ a
positive-definite Hermitian operator called the metric operator, playing a
role analogous to $\mathcal{CP}$.

While bound states of $\mathcal{PT}$-symmetric Hamiltonians are nowadays
well understood, many more questions remain open in the treatment of
scattering states: for instance, it has been shown in Ref.\cite{Jo07} that
even simple local potentials require introducing non-local metric operators
and non-standard boundary conditions with progressive and regressive waves
not only in the initial but also in the final state. Even if the latter
feature might be removed by an appropriate choice of the metric operator,
called quasi-local in Refs.\cite{Zn08a},\cite{Zn08b}, a satisfactory general
approach has not been formulated yet. This is why the majority of studies on
scattering by $\mathcal{PT}$-symmetric Hamiltonians has been made within the
framework of standard quantum mechanics, breaking unitarity of the
corresponding scattering matrices (see Ref.\cite{CDV07} and references
therein). Even at this effective level, $\mathcal{PT}$-symmetric potentials
are peculiar, in the sense that, depending on their parameters, they may
behave as absorptive for progressive waves and generative for regressive
ones (or viceversa), a property called handedness in Ref.\cite{Ah04}, or it
may happen that they are neither absorptive, nor generative, because the sum
of the square moduli of transmission and reflection coefficients may be
smaller than one, or greater than one in different intervals of incident
energy; they can even conserve unitarity when the asymptotic wave functions
are eigenstates of $\mathcal{PT}$: in this latter case they are necessarily
reflectionless\cite{CDV07}. As is known, the reflection of progressive
(left-to-right) and regressive (right-to-left) waves is quite asymmetric ($%
R_{L\rightarrow R}\neq R_{R\rightarrow L}$) already in the case of local
potentials, where the transmission is the same; in the case of non-local
potentials\cite{CV06}, the transmission is asymmetric, too ($T_{L\rightarrow
R}\neq T_{R\rightarrow L}$). Indeed, non-local potentials have more subtle $%
\mathcal{PT}$-transformation properties than local potentials, for which $%
\mathcal{T}$- invariance and Hermiticity requirements coincide\cite{CDV07}.

The scenario is even richer in relativistic quantum mechanics, where, again,
the majority of studies have been dedicated to bound states of $\mathcal{PT}$%
-symmetric potentials in the Klein-Gordon and Dirac equations in (1+1)
space-time dimensions. Limiting ourselves to the Dirac equation, of interest
to the present work, we may quote the pseudo-supersymmetric description\cite%
{SR05},\cite{SR06} of scalar or pseudo-scalar local potentials with exact,
or spontaneously broken $\mathcal{PT}$ \ symmetry, the $\mathcal{PT}$%
-symmetric version of the generalized Hulth\'{e}n vector potential\cite{ES05}
, the combinations of scalar (position-dependent mass) and vector potentials
of Refs.\cite{JD06},\cite{Ji07},\cite{JDL08},\cite{JD08}.

Making again an effective approach to scattering aspects, we have examined
in a recent work\cite{CV08} the Dirac equation with the time component of a
vector potential in the form of a $\mathcal{PT}$-symmetric square well: when
the real depth exceeds $2m$, with $m$ the particle mass, transmission
resonances at negative energies appear as the signature of spontaneous pair
production, but become weaker with increasing imaginary depth and negligible
beyond the critical value at which real bound states disappear.

In the present work , which extends the non-relativistic results of Refs.%
\cite{CDV07},\cite{CV06}, we consider a scalar and vector combination of
non-local separable potentials in the (1+1)-dimensional Dirac equation,
aimed in particular at the study of symmetries known in their
three-dimensional form as the spin and pseudo-spin symmetries, the latter
being experimentally observed in atomic nuclei . Numerical results will be
consistently obtained from the $\mathcal{PT}$-symmetric version of a
solvable potential originally proposed by Yamaguchi for the description of
bound and scattering states of the neutron-proton system.

Since this kind of potential has received until now moderate attention
within the framework of $\mathcal{PT}$-symmetric quantum mechanics, and, to
our knowledge, no attention at all in its relativistic version, we consider
it worthwhile to perform a detailed, albeit effective analysis by means of
the Green function method described in Section 2. The scattering matrix is
then studied in Section 3 and two non-relativistic limits for the particular
choices of the ratio of vector and scalar couplings \ corresponding to spin
and pseudo-spin symmetry are discussed in Section 4. Bound states are
studied in Section 5 and numerical results obtained with a kernel
corresponding to the Yamaguchi potential are discussed in Section 6. Section
7 is dedicated to conclusions and perspectives of future work.

\section{Green function approach}

Let us start with the (1+1)-dimensional Dirac equation with a
vector-plus-scalar non-local potential, written in units $\hbar =c=1$ 
\begin{eqnarray}
i\frac{\partial }{\partial t}\Psi \left( x,t\right) &=&\left( -i\alpha _{x}%
\frac{\partial }{\partial x}+\beta m\right) \Psi \left( x,t\right) +\left(
c_{S}\beta +c_{V}\right) \int_{-\infty }^{+\infty }dyK(x,y)\Psi \left(
y,t\right) \\
&\equiv &\left( \alpha _{x}p_{x}+\beta m\right) \Psi \left( x,t\right)
+\left( c_{S}\beta +c_{V}\right) \int_{-\infty }^{+\infty }dyK(x,y)\Psi
\left( y,t\right)  \notag \\
&\equiv &(H_{D}\Psi )\left( x,t\right) .  \label{Dirac_eq}
\end{eqnarray}

A stationary wave, $\Psi \left( x,t\right) =\Psi \left( x\right) e^{-iEt}$,
satisfies the equation 
\begin{equation}
(H_{D}\Psi )\left( x\right) =E\Psi \left( x\right) \;.  \label{Dirac_stat}
\end{equation}%
Here, $\alpha _{x}$ and $\beta $ are $2\times 2$ anticommuting Dirac
matrices with unit square, $\alpha _{x}^{2}=\beta ^{2}=\left( 
\begin{array}{cc}
1 & 0 \\ 
0 & 1%
\end{array}%
\right) \equiv 1_{2}$, which can be identified with two Pauli matrices: in
the present work we adopt the Dirac representation\cite{MS87} $\alpha
_{x}=\sigma _{x}\equiv \left( 
\begin{array}{cc}
0 & 1 \\ 
1 & 0%
\end{array}%
\right) $, $\beta =\sigma _{z}\equiv \left( 
\begin{array}{cc}
1 & 0 \\ 
0 & -1%
\end{array}%
\right) $, particularly suited to the study of the non-relativistic limit of
the model. $c_{S}$ and $c_{V}$ are the real strengths of the scalar
potential and of the time component of the vector potential, respectively,
with common $\mathcal{PT}$-symmetric kernel $K(x,y)=K^{\ast }\left(
-x,-y\right) $.

Here, as in our previous work\cite{CV08} on the one-dimensional Dirac
equation with a $\mathcal{PT}$-symmetric square well, we have the parity
operator $\mathcal{P}$ in the Dirac representation 
\begin{equation}
\mathcal{P}=e^{i\theta_{\mathcal{P}}}P_0\sigma_z\,  \label{P}
\end{equation}
where $P_0$ changes $x$ into $-x$ and $\theta_{\mathcal{P}}$ is an arbitrary
phase factor. In the same representation, the time reversal operator $%
\mathcal{T}$ reads 
\begin{equation}
\mathcal{T}=e^{i\theta_{\mathcal{T}}}\sigma_z\mathcal{K}\,  \label{T}
\end{equation}
where $\mathcal{K}$ performs complex conjugation and $\theta_{\mathcal{T}}$
is an arbitrary phase factor. With the convenient choice $\theta_{\mathcal{T}%
}=-\theta_{\mathcal{P}}$ the $\mathcal{PT}$ operator takes the form 
\begin{equation}
\mathcal{PT}=P_{0}\mathcal{K}\,  \label{PT}
\end{equation}
adopted also in non-relativistic quantum mechanics\cite{CDV07},\cite{CV06}.

It is worthwhile to point out that formula (\ref{Dirac_eq}) does not contain
the most general Hamiltonian: for instance, we might add a pseudo-scalar
interaction by extending the matrix of coupling strengths to $c_{S}\beta
+c_{V}+ic_{P}\alpha _{x}\beta $. The method of solution described in this
section could be applied even to the most general case, but we do not
consider it explicitly, because we are mainly interested in interaction
potentials that permit decoupling the two integro-differential equations
satisfied by the two components of $\Psi $, so as to obtain a clear
definition of their non-relativistic limits, as will be shown in detail in
Section 4.

In order to deal with a solvable model, we assume a separable kernel of the
form%
\begin{equation}
K(x,y)=g\left( x\right) e^{iax}h\left( y\right) e^{iby}\;,  \label{kernel}
\end{equation}%
where $a$ and $b$ are real numbers and the real functions $g$ and $h$ are
even functions of their arguments, $g\left( x\right) =g\left( -x\right) $
and $h\left( y\right) =h\left( -y\right) $, so as to assure $\mathcal{PT}$
invariance. When $g=h$ and $a=b=0$, the kernel becomes real symmetric and
coincides with that of Ref.\cite{CKN88}. When $g=h$ and $a=-b$ the kernel
becomes Hermitian, since in that case $K(x,y)=K^{\ast }\left( y,x\right) $.

Now we solve Eq. (\ref{Dirac_stat}) by means of the Green function method
already used for the one-dimensional Schr\"{o}dinger equation with the same
type of potential\cite{CDV07}\cite{CV06}. The Green function method had
already been used in the solution of a scalar-plus-vector real non-local
separable potential in Ref.\cite{CKN88} and of a pure vector potential in
Ref.\cite{DG92}.

Two linearly independent Green functions, $G_{+}\left( x,x^{\prime }\right) $
and $G_{-}\left( x,x^{\prime }\right) $, for the time-independent Dirac
equation (\ref{Dirac_stat}) are solutions to the equation%
\begin{equation}
\left( -i\alpha _{x}\frac{\partial }{\partial x}+\beta m-\left( E\pm
i\epsilon \right) \right) G_{\pm }(x,x^{\prime })=\delta \left( x-x^{\prime
}\right) \;,  \label{time_ind_Green}
\end{equation}%
where a small imaginary component $\epsilon $ $\left( >0\right) $ is added
to the energy, $E$, in order to remove the energy poles from the contour of
the complex integral defining $G_{\pm }(x,x^{\prime })$, as discussed in the
following part of this Section. $G_{+}\left( x,x^{\prime }\right) $ and $%
G_{-}\left( x,x^{\prime }\right) $ are related to the Laplace transform with
respect to time of the retarded and advanced component of the causal Green
function, respectively, as shown in Appendix A.

Eq. (\ref{time_ind_Green}) is easily solved in momentum space after
introducing the Fourier transforms $\widetilde{G}_{\pm }\left( q,q^{\prime
}\right) $%
\begin{equation}
G_{\pm }(x,x^{\prime })=\frac{1}{\left( 2\pi \right) ^{2}}\int_{-\infty
}^{+\infty }dqe^{iqx}\int_{-\infty }^{+\infty }dq^{\prime }e^{iq^{\prime
}x^{\prime }}\widetilde{G}_{\pm }\left( q,q^{\prime }\right)
\label{F_transf_G}
\end{equation}%
and the Fourier representation of the Dirac $\delta $ function%
\begin{equation}
\delta \left( x-x^{\prime }\right) =\frac{1}{2\pi }\int_{-\infty }^{+\infty
}dqe^{iq\left( x-x^{\prime }\right) }\;.  \label{F-transf_delta}
\end{equation}

After inserting formulae (\ref{F_transf_G}-\ref{F-transf_delta}) into Eq. (%
\ref{time_ind_Green}) we quickly obtain $\widetilde{G}_{\pm }\left(
q,q^{\prime }\right) $ in the form\bigskip 
\begin{equation}
\begin{array}{c}
\widetilde{G}_{\pm }\left( q,q^{\prime }\right) =\lim_{\epsilon \rightarrow
0+}2\pi \left( \alpha _{x}q+\beta m-E\mp i\epsilon \right) ^{-1}\delta
\left( q+q^{\prime }\right) \\ 
=2\pi \lim_{\epsilon \rightarrow 0+}\frac{\alpha _{x}q+\beta m+E\pm
i\epsilon }{q^{2}+m^{2}-\left( E\pm i\epsilon \right) ^{2}}\delta \left(
q+q^{\prime }\right)%
\end{array}%
\;.  \label{G_qq'}
\end{equation}

Therefore, we obtain for the Green functions in configuration space%
\begin{equation}
G_{\pm }\left( x,x^{\prime }\right) \equiv G_{\pm }\left( x-x^{\prime
}\right) =\frac{1}{2\pi }\lim_{\epsilon \rightarrow 0+}\int_{-\infty
}^{+\infty }dqe^{iq\left( x-x^{\prime }\right) }\frac{\alpha _{x}q+\beta
m+E\pm i\epsilon }{q^{2}+m^{2}-\left( E\pm i\epsilon \right) ^{2}}\;,
\label{G_xx'}
\end{equation}%
which can be easily computed by the method of residues. Let us start with $%
G_{+}\left( x-x^{\prime }\right) $: after defining $k^{2}=E^{2}-m^{2}$, we
observe that the integrand has two simple poles at $q_{1}=-k-i\epsilon
^{\prime }$ and $q_{2}=+k+i\epsilon ^{\prime }$, where $\epsilon ^{\prime
}=\epsilon E/k$. For $x-x^{\prime }\geq 0$ the integration contour is closed
in the upper $q$ half-plane, including the pole at $q=q_{2}$, while for $%
x-x^{\prime }<0$ the contour is closed in the lower $q$ half-plane,
including the pole at $q=q_{1}$ with a global $-$ sign, because the
integration is done in the clockwise direction. The result is%
\begin{equation}
\begin{array}{c}
G_{+}\left( x-x^{\prime }\right) =\frac{i}{2k}\left[ \theta \left(
x-x^{\prime }\right) e^{ik\left( x-x^{\prime }\right) }\left( \alpha
_{x}k+\beta m+E\right) \right. \\ 
\left. +\theta \left( x^{\prime }-x\right) e^{-ik\left( x-x^{\prime }\right)
}\left( -\alpha _{x}k+\beta m+E\right) \right] \\ 
=\frac{i}{2k}e^{ik\left\vert x-x^{\prime }\right\vert }\left( k\alpha
_{x}sgn\left( x-x^{\prime }\right) +\beta m+E\right) \;,%
\end{array}
\label{Gplus}
\end{equation}%
in agreement with Ref.\cite{CKN88}. In the same way we compute $G_{-}\left(
x-x^{\prime }\right) $, after observing that for $x-x^{\prime }\geq 0$ the
integration contour in the upper $q$ half-plane now includes a pole at $%
q_{3}=-k+i\epsilon ^{\prime }$ while, for $x-x^{\prime }<0$, the contour is
closed in the clockwise direction in the lower $q$ half-plane around a pole
at $q_{4}=+k-i\epsilon ^{\prime }$. The result is 
\begin{equation}
\begin{array}{c}
G_{-}\left( x-x^{\prime }\right) =-\frac{i}{2k}\left[ \theta \left(
x-x^{\prime }\right) e^{-ik\left( x-x^{\prime }\right) }\left( -\alpha
_{x}k+\beta m+E\right) \right. \\ 
\left. +\theta \left( x^{\prime }-x\right) e^{ik\left( x-x^{\prime }\right)
}\left( \alpha _{x}k+\beta m+E\right) \right] \\ 
=-\frac{i}{2k}e^{-ik\left\vert x-x^{\prime }\right\vert }\left( -k\alpha
_{x}sgn\left( x-x^{\prime }\right) +\beta m+E\right) \;.%
\end{array}
\label{Gminus}
\end{equation}

Summing up%
\begin{equation}
G_{\pm }\left( x-x^{\prime }\right) =\pm \frac{i}{2k}e^{\pm ik\left\vert
x-x^{\prime }\right\vert }\left( \pm k\alpha _{x}sgn\left( x-x^{\prime
}\right) +\beta m+E\right) \;.  \label{Gpm}
\end{equation}

It is immediate to check that%
\begin{equation*}
G_{-}\left( x-x^{\prime }\right) =\mathcal{PT}G_{+}\left( x-x^{\prime
}\right) \left( \mathcal{PT}\right) ^{-1}\;.
\end{equation*}

\section{Scattering matrix}

By exploiting the results of the preceding section, we can define two
linearly independent solutions to Eq. (\ref{Dirac_stat}), $\Psi _{+}\left(
x\right) $ and $\Psi _{-}\left( x\right) $, in the implicit form%
\begin{equation}
\begin{array}{c}
\Psi _{\pm }\left( x\right) =\Psi _{free}^{\pm }\left( x\right)
-\int_{-\infty }^{+\infty }dx^{\prime }G_{\pm }\left( x-x^{\prime }\right)
\left( c_{S}\beta +c_{V}\right) \int_{-\infty }^{+\infty }dyK\left(
x^{\prime },y\right) \Psi _{\pm }\left( y\right) \\ 
=\Psi _{free}^{\pm }\left( x\right) -\int_{-\infty }^{+\infty }dx^{\prime
}G_{\pm }\left( x-x^{\prime }\right) \left( c_{S}\beta +c_{V}\right) g\left(
x^{\prime }\right) e^{iax^{\prime }}\int_{-\infty }^{+\infty }dyh\left(
y\right) e^{iby}\Psi _{\pm }\left( y\right) \;.%
\end{array}
\label{Psi_pm}
\end{equation}

In Eq. (\ref{Psi_pm}), $\Psi _{free}^{\pm }\left( x\right) $ is the general
solution to the Dirac equation for a free particle, conveniently written in
the matrix notation of Ref.\cite{MS87}%
\begin{equation}
\Psi _{free}^{\pm }\left( x\right) =\left( 
\begin{array}{cc}
e^{ikx} & e^{-ikx} \\ 
\lambda e^{ikx} & -\lambda e^{-ikx}%
\end{array}%
\right) \cdot \left( 
\begin{array}{c}
A_{\pm } \\ 
B_{\pm }%
\end{array}%
\right) =\left( 
\begin{array}{c}
A_{\pm }e^{ikx}+B_{\pm }e^{-ikx} \\ 
\lambda A_{\pm }e^{ikx}-\lambda B_{\pm }e^{-ikx}%
\end{array}%
\right) \;,  \label{Psi_free}
\end{equation}%
where $\lambda \equiv k/\left( E+m\right) =$ $\sqrt{\left( E-m\right)
/\left( E+m\right) }$ and $A_{\pm }$ and $B_{\pm }$ are arbitrary constants.
It is worthwhile to point out that $G_{\pm }\left( x-x^{\prime }\right) $
and $c_{S}\beta +c_{V}$ are non-commuting $2\times 2$ matrices: therefore,
their order is not arbitrary.

After defining $I_{\pm }\equiv \int_{-\infty }^{+\infty }dyh\left( y\right)
e^{iby}\Psi _{\pm }\left( y\right) $, we multiply both sides of Eq. (\ref%
{Psi_pm}) by $h\left( x\right) e^{ibx}$ and integrate them over $x$ from $%
-\infty $ to $+\infty $. Remembering that $\widetilde{f}\left( q\right)
=\int_{-\infty }^{+\infty }dxe^{-iqx}f\left( x\right) $ is the Fourier
transform of $f\left( x\right) $ and observing that $f\left( x\right)
=f\left( -x\right) $ implies $\widetilde{f}\left( q\right) =\widetilde{f}%
\left( -q\right) $, we promptly obtain 
\begin{equation}
\begin{array}{c}
I_{\pm }=\left( 
\begin{array}{c}
A_{\pm }\widetilde{h}\left( k+b\right) +B_{\pm }\widetilde{h}\left(
k-b\right)  \\ 
\lambda A_{\pm }\widetilde{h}\left( k+b\right) -\lambda B_{\pm }\widetilde{h}%
\left( k-b\right) 
\end{array}%
\right)  \\ 
-\int_{-\infty }^{+\infty }dxh\left( x\right) e^{ibx}\int_{-\infty
}^{+\infty }dx^{\prime }G_{\pm }\left( x-x^{\prime }\right) \left(
c_{S}\beta +c_{V}\right) g\left( x^{\prime }\right) e^{iax^{\prime }}I_{\pm }
\\ 
=\left( 
\begin{array}{c}
A_{\pm }\widetilde{h}\left( k+b\right) +B_{\pm }\widetilde{h}\left(
k-b\right)  \\ 
\lambda A_{\pm }\widetilde{h}\left( k+b\right) -\lambda B_{\pm }\widetilde{h}%
\left( k-b\right) 
\end{array}%
\right) -N_{\pm }\left( c_{S}\beta +c_{V}\right) I_{\pm }\;,%
\end{array}
\label{Ipm_eq}
\end{equation}%
where $N_{\pm }\equiv \int_{-\infty }^{+\infty }dxh\left( x\right)
e^{ibx}\int_{-\infty }^{+\infty }dx^{\prime }G_{\pm }\left( x-x^{\prime
}\right) g\left( x^{\prime }\right) e^{iax^{\prime }}$. Therefore, the
spinor $I_{\pm }$ is explicitly given by the relation 
\begin{equation}
I_{\pm }=\left( 1_{2}+N_{\pm }\left( c_{S}\beta +c_{V}\right) \right)
^{-1}\cdot \left( 
\begin{array}{c}
A_{\pm }\widetilde{h}\left( k+b\right) +B_{\pm }\widetilde{h}\left(
k-b\right)  \\ 
\lambda A_{\pm }\widetilde{h}\left( k+b\right) -\lambda B_{\pm }\widetilde{h}%
\left( k-b\right) 
\end{array}%
\right) \;,  \label{Ipm}
\end{equation}%
once we have determined the $2\times 2$ matrix $N_{\pm }$, which, according
to formula (\ref{Gpm}), is conveniently rewritten as%
\begin{equation}
\begin{array}{c}
N_{\pm }=\pm \frac{i}{2k}\int_{-\infty }^{+\infty }dxh\left( x\right)
e^{ibx}\int_{-\infty }^{+\infty }dx^{\prime }g\left( x^{\prime }\right)
e^{iax^{\prime }}\left[ e^{\pm ik\left( x-x^{\prime }\right) }\left( \pm
\alpha _{x}k+\beta m+E\right) \theta \left( x-x^{\prime }\right) \right.  \\ 
\left. +e^{\mp ik\left( x-x^{\prime }\right) }\left( \mp \alpha _{x}k+\beta
m+E\right) \theta \left( x^{\prime }-x\right) \right]  \\ 
=\pm \frac{i}{2k}\left[ N_{\pm }^{\left( 1\right) }\left( \pm \alpha
_{x}k+\beta m+E\right) +N_{\pm }^{\left( 2\right) }\left( \mp \alpha
_{x}k+\beta m+E\right) \right] \;,%
\end{array}
\label{Npm}
\end{equation}%
where%
\begin{equation}
N_{\pm }^{\left( 1\right) }\left( a,b,k\right) \equiv \int_{-\infty
}^{+\infty }dxh\left( x\right) e^{ibx}\int_{-\infty }^{+\infty }dx^{\prime
}g\left( x^{\prime }\right) e^{iax^{\prime }}e^{\pm ik\left( x-x^{\prime
}\right) }\theta \left( x-x^{\prime }\right)   \label{N1pm}
\end{equation}%
and%
\begin{equation}
N_{\pm }^{\left( 2\right) }\left( a,b,k\right) \equiv \int_{-\infty
}^{+\infty }dxh\left( x\right) e^{ibx}\int_{-\infty }^{+\infty }dx^{\prime
}g\left( x^{\prime }\right) e^{iax^{\prime }}e^{\mp ik\left( x-x^{\prime
}\right) }\theta \left( x^{\prime }-x\right) \;.  \label{N2pm}
\end{equation}

It is worthwhile to point out the following symmetry relation%
\begin{equation}
N_{+}^{\left( j\right) }\left( -a,-b,k\right) =\left( N_{-}^{\left( j\right)
}\left( a,b,k\right) \right) ^{\ast }\;\left( j=1,2\right) \;.  \label{Sym1}
\end{equation}

After introducing the linear combinations 
\begin{equation}
S_{\pm }\left( a,b,k\right) \equiv N_{\pm }^{\left( 1\right) }\left(
a,b,k\right) +N_{\pm }^{\left( 2\right) }\left( a,b,k\right) \;,\quad D_{\pm
}\left( a,b,k\right) =N_{\pm }^{\left( 1\right) }\left( a,b,k\right) -N_{\pm
}^{\left( 2\right) }\left( a,b,k\right) \;,  \label{S_D}
\end{equation}%
with symmetry relations%
\begin{equation}
S_{+}\left( -a,-b,k\right) =\left( S_{-}\left( a,b,k\right) \right) ^{\ast
},\;D_{+}\left( -a,-b,k\right) =\left( D_{-}\left( a,b,k\right) \right)
^{\ast },  \label{Sym2}
\end{equation}%
$N_{\pm }$ becomes%
\begin{equation}
N_{\pm }=\frac{i}{2k}\left[ D_{\pm }\alpha _{x}k\pm S_{\pm }\left( \beta
m+E\right) \right] \;.  \label{Npm_bis}
\end{equation}

We now specialize to the Dirac representation, already introduced in Section
2, $\alpha _{x}=\sigma _{x}$, $\beta =\sigma _{z}$. After some simple
algebra, we obtain%
\begin{equation}
N_{\pm }=\left( 
\begin{array}{cc}
\pm i\frac{S_{\pm }}{2\lambda } & i\frac{D_{\pm }}{2} \\ 
i\frac{D_{\pm }}{2} & \pm i\frac{S_{\pm }\lambda }{2}%
\end{array}%
\right) \;,  \label{Npm_ter}
\end{equation}

\begin{equation}
\begin{array}{c}
M_{\pm }\equiv 1_{2}+N_{\pm }\left( c_{S}\beta +c_{V}\right) \\ 
=\left( 
\begin{array}{cc}
1\pm \frac{i}{2}\frac{S_{\pm }}{\lambda }\left( c_{V}+c_{S}\right) & \frac{i%
}{2}D_{\pm }\left( c_{V}-c_{S}\right) \\ 
\frac{i}{2}D_{\pm }\left( c_{V}+c_{S}\right) & 1\pm \frac{i}{2}\lambda
S_{\pm }\left( c_{V}-c_{S}\right)%
\end{array}%
\right)%
\end{array}%
\;.  \label{Mpm}
\end{equation}

In order to compute $I_{\pm }$ from formula (\ref{Ipm}), we need the inverse
of $M_{\pm }$%
\begin{equation}
M_{\pm }^{-1}=\frac{1}{\det M_{\pm }}\left( 
\begin{array}{cc}
1\pm \frac{i}{2}\lambda S_{\pm }\left( c_{V}-c_{S}\right) & -\frac{i}{2}%
D_{\pm }\left( c_{V}-c_{S}\right) \\ 
-\frac{i}{2}D_{\pm }\left( c_{V}+c_{S}\right) & 1\pm \frac{i}{2}\frac{S_{\pm
}}{\lambda }\left( c_{V}+c_{S}\right)%
\end{array}%
\right) \;,  \label{Mpm_m1}
\end{equation}%
with%
\begin{equation}
\begin{array}{c}
\det M_{\pm }=\left( 1\pm \frac{i}{2}\lambda S_{\pm }\left(
c_{V}-c_{S}\right) \right) \left( 1\pm \frac{i}{2}\frac{S_{\pm }}{\lambda }%
\left( c_{V}+c_{S}\right) \right) +\frac{D_{\pm }^{2}}{4}\left(
c_{V}^{2}-c_{S}^{2}\right) \\ 
=1\pm \frac{i}{2}S_{\pm }\left( \lambda \left( c_{V}-c_{S}\right) +\frac{1}{%
\lambda }\left( c_{V}+c_{S}\right) \right) +\frac{c_{V}^{2}-c_{S}^{2}}{4}%
\left( D_{\pm }^{2}-S_{\pm }^{2}\right) \\ 
=1\pm i\frac{S_{\pm }}{k}\left( c_{V}E+c_{S}m\right) +\frac{%
c_{V}^{2}-c_{S}^{2}}{4}\left( D_{\pm }^{2}-S_{\pm }^{2}\right) \;.%
\end{array}
\label{detMpm}
\end{equation}

Note that, as a consequence of relations (\ref{Sym1}-\ref{Sym2}),%
\begin{equation}
\det M_{-}\left( -a,-b,k\right) =\left( \det M_{+}\left( a,b,k\right)
\right) ^{\ast }\;.  \label{Sym3}
\end{equation}

We are now in a position to express the asymptotic forms of the wave
functions $\Psi _{\pm }\left( x\right) $ and the transmission and reflection
coefficients for progressive and regressive waves in terms of known
quantities. For the sake of clarity, let us consider $\Psi _{+}\left(
x\right) $ and $\Psi _{-}\left( x\right) $ separately.

In order to determine the asymptotic behaviour of $\Psi _{+}\left( x\right) $%
, we observe that%
\begin{equation}
\lim_{x\rightarrow \pm \infty }G_{+}\left( x-x^{\prime }\right) =\frac{i}{2}%
e^{\pm ik\left( x-x^{\prime }\right) }\left( 
\begin{array}{cc}
\frac{1}{\lambda } & \pm 1 \\ 
\pm 1 & \lambda%
\end{array}%
\right) \;.
\end{equation}

Therefore, in particular%
\begin{equation}
\begin{array}{c}
\lim_{x\rightarrow +\infty }\Psi _{+}\left( x\right) =A_{+}\left( 
\begin{array}{c}
1 \\ 
\lambda%
\end{array}%
\right) e^{ikx}+B_{+}\left( 
\begin{array}{c}
1 \\ 
-\lambda%
\end{array}%
\right) e^{-ikx} \\ 
-\frac{i}{2}\int_{-\infty }^{+\infty }dx^{\prime }e^{ik\left( x-x^{\prime
}\right) }e^{iax^{\prime }}g\left( x^{\prime }\right) \left( 
\begin{array}{cc}
\frac{1}{\lambda } & 1 \\ 
1 & \lambda%
\end{array}%
\right) \cdot \left( 
\begin{array}{cc}
c_{V}+c_{S} & 0 \\ 
0 & c_{V}-c_{S}%
\end{array}%
\right) \cdot I_{+} \\ 
=A_{+}\left( 
\begin{array}{c}
1 \\ 
\lambda%
\end{array}%
\right) e^{ikx}+B_{+}\left( 
\begin{array}{c}
1 \\ 
-\lambda%
\end{array}%
\right) e^{-ikx} \\ 
-\frac{i}{2}\widetilde{g}\left( a-k\right) \left( 
\begin{array}{cc}
\frac{c_{V}+c_{S}}{\lambda } & c_{V}-c_{S} \\ 
c_{V}+c_{S} & \lambda \left( c_{V}-c_{S}\right)%
\end{array}%
\right) \cdot I_{+}e^{ikx}\;.%
\end{array}%
\end{equation}

If we impose the condition that $\Psi _{+}\left( x\right) $ is a progressive
wave, travelling from left to right ($L\rightarrow R$), we can put $A_{+}=1$
and $B_{+}=0$ in the preceding equation. After deriving from formula (\ref%
{Ipm}) the explicit form of $I_{+}$ 
\begin{equation}
I_{+}=\frac{\widetilde{h}\left( k+b\right) }{\det M_{+}}\left( 
\begin{array}{c}
1+\frac{i}{2}\lambda S_{+}\left( c_{V}-c_{S}\right) -\frac{i}{2}\lambda
D_{+}\left( c_{V}-c_{S}\right) \\ 
-\frac{i}{2}D_{+}\left( c_{V}+c_{S}\right) +\lambda +\frac{i}{2}S_{+}\left(
c_{V}+c_{S}\right)%
\end{array}%
\right) \;,  \label{i_p}
\end{equation}%
the above limit can be rewritten after some algebra in the form%
\begin{equation}
\begin{array}{c}
\lim_{x\rightarrow +\infty }\Psi _{+}\left( x\right) =\left( 
\begin{array}{c}
1 \\ 
\lambda%
\end{array}%
\right) e^{ikx}\left[ 1-\frac{i}{2}\widetilde{g}\left( a-k\right) \widetilde{%
h}\left( k+b\right) \times \right. \\ 
\left. \frac{\frac{2}{k}\left( c_{V}E+c_{S}m\right) +i\left(
c_{V}^{2}-c_{S}^{2}\right) \left( S_{+}-D_{+}\right) }{1+\frac{S_{+}}{k}%
\left( c_{V}E+c_{S}m\right) +\frac{1}{4}\left( c_{V}^{2}-c_{S}^{2}\right)
\left( D_{+}^{2}-S_{+}^{2}\right) }\right]%
\end{array}
\label{Psip_pinf}
\end{equation}%
allowing us to determine the transmission coefficient, $T_{L\rightarrow R}$,
since we must have%
\begin{equation}
\lim_{x\rightarrow +\infty }\Psi _{+}\left( x\right) =T_{L\rightarrow
R}\left( 
\begin{array}{c}
1 \\ 
\lambda%
\end{array}%
\right) e^{ikx}\;.  \label{Psip_pinf1}
\end{equation}

From comparison of the r.h.s. of Eqs (\ref{Psip_pinf}) and (\ref{Psip_pinf1}%
) we obtain 
\begin{equation}
T_{L\rightarrow R}=1-\frac{i}{2}\widetilde{g}\left( a-k\right) \widetilde{h}%
\left( k+b\right) \frac{\frac{2}{k}\left( c_{V}E+c_{S}m\right) +i\left(
c_{V}^{2}-c_{S}^{2}\right) \left( S_{+}-D_{+}\right) }{1+i\frac{S_{+}}{k}%
\left( c_{V}E+c_{S}m\right) +\frac{1}{4}\left( c_{V}^{2}-c_{S}^{2}\right)
\left( D_{+}^{2}-S_{+}^{2}\right) }\;.  \label{T_LR}
\end{equation}

In the same way we can compute the reflection coefficient, $R_{L\rightarrow
R}$, starting from%
\begin{equation}
\begin{array}{c}
\lim_{x\rightarrow -\infty }\Psi _{+}\left( x\right) =\left( 
\begin{array}{c}
1 \\ 
\lambda%
\end{array}%
\right) e^{ikx} \\ 
-\frac{i}{2}e^{-ikx}\int_{-\infty }^{+\infty }dx^{\prime }e^{i\left(
k+a\right) x^{\prime }}g\left( x^{\prime }\right) \left( 
\begin{array}{cc}
\frac{1}{\lambda } & -1 \\ 
-1 & \lambda%
\end{array}%
\right) \cdot \left( 
\begin{array}{cc}
c_{V}+c_{S} & 0 \\ 
0 & c_{V}-c_{S}%
\end{array}%
\right) \cdot I_{+} \\ 
=\left( 
\begin{array}{c}
1 \\ 
\lambda%
\end{array}%
\right) e^{ikx}-\frac{i}{2}e^{-ikx}\widetilde{g}\left( k+a\right) \left( 
\begin{array}{cc}
\frac{c_{V}+c_{S}}{\lambda } & c_{S}-c_{V} \\ 
-\left( c_{V}+c_{S}\right) & \lambda \left( c_{V}-c_{S}\right)%
\end{array}%
\right) \cdot I_{+}\;.%
\end{array}%
\end{equation}

Using again formula (\ref{i_p}) for $I_{+}$, we obtain after some simple
algebra%
\begin{equation}
\lim_{x\rightarrow -\infty }\Psi _{+}\left( x\right) =\left( 
\begin{array}{c}
1 \\ 
\lambda%
\end{array}%
\right) e^{ikx}-\frac{i}{2}\frac{\widetilde{g}\left( k+a\right) \widetilde{h}%
\left( k+b\right) }{\det M_{+}}\left( \frac{c_{V}+c_{S}}{\lambda }+\lambda
\left( c_{S}-c_{V}\right) \right) \left( 
\begin{array}{c}
1 \\ 
-\lambda%
\end{array}%
\right) e^{-ikx}\;,  \label{Psip_minf}
\end{equation}%
where $\det M_{+}$ is given by formula (\ref{detMpm}). On the other hand, we
must have%
\begin{equation}
\lim_{x\rightarrow -\infty }\Psi _{+}\left( x\right) =\left( 
\begin{array}{c}
1 \\ 
\lambda%
\end{array}%
\right) e^{ikx}+R_{L\rightarrow R}\left( 
\begin{array}{c}
1 \\ 
-\lambda%
\end{array}%
\right) e^{-ikx}\;.  \label{Psip_minf1}
\end{equation}

From formulae (\ref{Psip_minf}) and (\ref{Psip_minf1}) we promptly obtain%
\begin{equation}
R_{L\rightarrow R}=-\frac{i}{k}\widetilde{g}\left( k+a\right) \widetilde{h}%
\left( k+b\right) \frac{c_{V}m+c_{S}E}{1+i\frac{S_{+}}{k}\left(
c_{V}E+c_{S}m\right) +\frac{1}{4}\left( c_{V}^{2}-c_{S}^{2}\right) \left(
D_{+}^{2}-S_{+}^{2}\right) }\;.  \label{R_LR}
\end{equation}

In order to compare our results with those of Ref.\cite{CKN88} for a real
symmetric kernel, with $g\left( x\right) =h\left( x\right) \equiv v\left(
x\right) $ and $a=b=0$, we observe that, in this limit, $D_{+}=0$ and $%
S_{+}\equiv J=\int_{-\infty }^{+\infty }dx\int_{-\infty }^{+\infty
}dx^{\prime }e^{ik\left\vert x-x^{\prime }\right\vert }v\left( x\right)
v\left( x^{\prime }\right) $. $J$ is promptly expressed in terms of the
Fourier transform of $v\left( x\right) $, $\widetilde{v}\left( k\right)
\equiv \int_{-\infty }^{+\infty }dxv\left( x\right) e^{-ikx}$. In fact%
\begin{equation}
J=J_{R}+iJ_{I}=\int_{-\infty }^{+\infty }dx\int_{-\infty }^{+\infty
}dx^{\prime }v\left( x\right) v\left( x^{\prime }\right) \left[ \cos
k\left\vert x-x^{\prime }\right\vert +i\sin k\left\vert x-x^{\prime
}\right\vert \right] \;,
\end{equation}%
where%
\begin{equation}
\begin{array}{c}
J_{R}=\int_{-\infty }^{+\infty }dx\int_{-\infty }^{+\infty }dx^{\prime
}v\left( x\right) v\left( x^{\prime }\right) \cos k(x-x^{\prime }) \\ 
=\int_{-\infty }^{+\infty }dx\int_{-\infty }^{+\infty }dx^{\prime }v\left(
x\right) v\left( x^{\prime }\right) e^{ik\left( x-x^{\prime }\right) }=%
\widetilde{v}\left( k\right) \widetilde{v}\left( -k\right) =\left( 
\widetilde{v}\left( k\right) \right) ^{2}%
\end{array}%
\end{equation}%
and%
\begin{equation}
iJ_{I}=J-J_{R}=J-\left( \widetilde{v}\left( k\right) \right) ^{2}\;.
\end{equation}

Thus, in the same limit, we obtain $S_{+}-\widetilde{g}\left( k\right) 
\widetilde{h}\left( k\right) =J-\left( \widetilde{v}\left( k\right) \right)
^{2}=J-J_{R}=iJ_{I}$ and $S_{+}^{2}-2\widetilde{g}\left( k\right) \widetilde{%
h}\left( k\right) S_{+}=J\left( J-2J_{R}\right) =-JJ^{\ast }=-\left\vert
J\right\vert ^{2}$. Therefore, our $\left\vert T_{L\rightarrow R}\right\vert
^{2}$, from formula (\ref{T_LR}), coincides with formula (13) of Ref.\cite%
{CKN88} and our $\left\vert R_{L\rightarrow R}\right\vert ^{2}$, from
formula (\ref{R_LR}), with formula (14) of the same reference, as expected.

Let us now consider the second Green function, $G_{-}\left( x-x^{\prime
}\right) $, whose asymptotic behaviour is%
\begin{equation}
\lim_{x\rightarrow \pm \infty }G_{-}\left( x-x^{\prime }\right) =-\frac{i}{2}%
e^{\mp ik\left( x-x^{\prime }\right) }\left( 
\begin{array}{cc}
\frac{1}{\lambda } & \mp 1 \\ 
\mp 1 & \lambda%
\end{array}%
\right) \;.
\end{equation}

Therefore, in particular%
\begin{equation}
\begin{array}{c}
\lim_{x\rightarrow -\infty }\Psi _{-}\left( x\right) =A_{-}\left( 
\begin{array}{c}
1 \\ 
\lambda%
\end{array}%
\right) e^{ikx}+B_{-}\left( 
\begin{array}{c}
1 \\ 
-\lambda%
\end{array}%
\right) e^{-ikx} \\ 
+\frac{i}{2}\int_{-\infty }^{+\infty }dx^{\prime }e^{ik\left( x-x^{\prime
}\right) }e^{iax^{\prime }}g\left( x^{\prime }\right) \left( 
\begin{array}{cc}
\frac{1}{\lambda } & 1 \\ 
1 & \lambda%
\end{array}%
\right) \cdot \left( 
\begin{array}{cc}
c_{S}+c_{V} & 0 \\ 
0 & c_{V}-c_{S}%
\end{array}%
\right) \cdot I_{-} \\ 
=\left[ A_{-}\left( 
\begin{array}{c}
1 \\ 
\lambda%
\end{array}%
\right) +\frac{i}{2}\widetilde{g}\left( a-k\right) \left( 
\begin{array}{cc}
\frac{c_{S}+c_{V}}{\lambda } & c_{V}-c_{S} \\ 
c_{S}+c_{V} & \lambda \left( c_{V}-c_{S}\right)%
\end{array}%
\right) \cdot I_{-}\right] e^{ikx}+B_{-}\left( 
\begin{array}{c}
1 \\ 
-\lambda%
\end{array}%
\right) e^{-ikx}\;.%
\end{array}
\label{Psim_minf}
\end{equation}

We can impose the condition that $\Psi _{-}\left( x\right) $ is a regressive
wave, travelling from right to left ($R\rightarrow L$), so that%
\begin{equation}
\lim_{x\rightarrow -\infty }\Psi _{-}\left( x\right) =T_{R\rightarrow
L}\left( 
\begin{array}{c}
1 \\ 
-\lambda%
\end{array}%
\right) e^{-ikx}\;.  \label{Psim_minf1}
\end{equation}

Comparison of formulae (\ref{Psim_minf}) and (\ref{Psim_minf1}) yields%
\begin{equation}
\left\{ 
\begin{array}{c}
B_{-}=T_{R\rightarrow L} \\ 
A_{-}\left( 
\begin{array}{c}
1 \\ 
\lambda%
\end{array}%
\right) +\frac{i}{2}\widetilde{g}\left( a-k\right) \left( 
\begin{array}{cc}
\frac{c_{S}+c_{V}}{\lambda } & c_{V}-c_{S} \\ 
c_{S}+c_{V} & \lambda \left( c_{V}-c_{S}\right)%
\end{array}%
\right) \cdot I_{-}=\left( 
\begin{array}{c}
0 \\ 
0%
\end{array}%
\right)%
\end{array}%
\right. \;.  \label{Eqn_arr1}
\end{equation}

In the same way%
\begin{equation}
\begin{array}{c}
\lim_{x\rightarrow +\infty }\Psi _{-}\left( x\right) =A_{-}\left( 
\begin{array}{c}
1 \\ 
\lambda%
\end{array}%
\right) e^{ikx}+B_{-}\left( 
\begin{array}{c}
1 \\ 
-\lambda%
\end{array}%
\right) e^{-ikx} \\ 
+\frac{i}{2}\int_{-\infty }^{+\infty }dx^{\prime }e^{-ik\left( x-x^{\prime
}\right) }e^{iax^{\prime }}g\left( x^{\prime }\right) \left( 
\begin{array}{cc}
\frac{1}{\lambda } & -1 \\ 
-1 & \lambda%
\end{array}%
\right) \cdot \left( 
\begin{array}{cc}
c_{S}+c_{V} & 0 \\ 
0 & c_{V}-c_{S}%
\end{array}%
\right) \cdot I_{-} \\ 
=A_{-}\left( 
\begin{array}{c}
1 \\ 
\lambda%
\end{array}%
\right) e^{ikx}+\left[ B_{-}\left( 
\begin{array}{c}
1 \\ 
-\lambda%
\end{array}%
\right) +\frac{i}{2}\widetilde{g}\left( a+k\right) \left( 
\begin{array}{cc}
\frac{c_{S}+c_{V}}{\lambda } & c_{S}-c_{V} \\ 
-(c_{S}+c_{V}) & \lambda \left( c_{V}-c_{S}\right)%
\end{array}%
\right) \cdot I_{-}\right] e^{-ikx}\;.%
\end{array}
\label{Psim_pinf}
\end{equation}

Since we know that%
\begin{equation}
\lim_{x\rightarrow +\infty }\Psi _{-}\left( x\right) =\left( 
\begin{array}{c}
1 \\ 
-\lambda%
\end{array}%
\right) e^{-ikx}+R_{R\rightarrow L}\left( 
\begin{array}{c}
1 \\ 
\lambda%
\end{array}%
\right) e^{ikx}\;,  \label{Psim_pinf1}
\end{equation}%
we obtain%
\begin{equation}
\left\{ 
\begin{array}{c}
A_{\_}=R_{R\rightarrow L} \\ 
B_{-}\left( 
\begin{array}{c}
1 \\ 
-\lambda%
\end{array}%
\right) +\frac{i}{2}\widetilde{g}\left( a+k\right) \left( 
\begin{array}{cc}
\frac{c_{S}+c_{V}}{\lambda } & c_{S}-c_{V} \\ 
-(c_{S}+c_{V}) & \lambda \left( c_{V}-c_{S}\right)%
\end{array}%
\right) \cdot I_{-}=\left( 
\begin{array}{c}
1 \\ 
-\lambda%
\end{array}%
\right)%
\end{array}%
\right. \;.  \label{Eqn_arr2}
\end{equation}

Remembering the expression of $I_{-}$ from formulae (\ref{Ipm}-\ref{Mpm_m1}-%
\ref{detMpm}), \ and rewriting it more compactly as%
\begin{equation}
I_{-}=\frac{1}{\det M_{-}}\left( 
\begin{array}{cc}
1-\frac{i}{2}\lambda S_{-}\left( c_{V}-c_{S}\right) & -\frac{i}{2}\lambda
D_{-}\left( c_{V}-c_{S}\right) \\ 
-\frac{i}{2}D_{-}\left( c_{V}+c_{S}\right) & 1-\frac{i}{2}\frac{S_{-}}{%
\lambda }\left( c_{V}+c_{S}\right)%
\end{array}%
\right) \cdot \left( 
\begin{array}{c}
\mathfrak{S} \\ 
\lambda \mathfrak{D}%
\end{array}%
\right) \;,
\end{equation}%
with $\mathfrak{S\equiv }A_{\_}\widetilde{h}\left( k+b\right) +B_{-}%
\widetilde{h}\left( k-b\right) =R_{R\rightarrow L}\widetilde{h}\left(
k+b\right) +T_{R\rightarrow L}\widetilde{h}\left( k-b\right) $ and $%
\mathfrak{D\equiv }R_{R\rightarrow L}\widetilde{h}\left( k+b\right)
-T_{R\rightarrow L}\widetilde{h}\left( k-b\right) $, \ we obtain from Eqs. (%
\ref{Eqn_arr1}-\ref{Eqn_arr2}) a system of two linear equations in the
unknowns $\mathfrak{S}$ and $\mathfrak{D}$%
\begin{equation}
\left\{ 
\begin{array}{c}
A_{-}+\frac{i}{2}\frac{\widetilde{g}\left( a-k\right) }{\det M_{-}}\mathfrak{%
P}_{+}=0 \\ 
B_{-}+\frac{i}{2}\frac{\widetilde{g}\left( a+k\right) }{\det M_{-}}\mathfrak{%
P}_{-}=1%
\end{array}%
\right. \;\;^{\prime }  \label{Eqn_arr3}
\end{equation}%
with%
\begin{equation}
\begin{array}{c}
\mathfrak{P}_{\pm }=\left[ \frac{c_{S}+c_{V}}{\lambda }-\frac{i}{2}\left(
c_{V}^{2}-c_{S}^{2}\right) \left( S_{-}\pm D_{-}\right) \right] \mathfrak{%
S\pm }\left[ \lambda \left( c_{V}-c_{S}\right) -\frac{i}{2}\left(
c_{V}^{2}-c_{S}^{2}\right) \left( S_{-}\pm D_{-}\right) \right] \mathfrak{D}
\\ 
\equiv \mathfrak{P}_{\pm }^{\left( \mathfrak{S}\right) }\mathfrak{S+P}_{\pm
}^{\left( \mathfrak{D}\right) }\mathfrak{D\;,}%
\end{array}
\label{Pmp}
\end{equation}%
where%
\begin{equation}
\begin{array}{c}
\mathfrak{P}_{\pm }^{\left( \mathfrak{S}\right) }\equiv \left[ \frac{%
c_{S}+c_{V}}{\lambda }-\frac{i}{2}\left( c_{V}^{2}-c_{S}^{2}\right) \left(
S_{-}\pm D_{-}\right) \right] \;, \\ 
\mathfrak{P}_{\pm }^{\left( \mathfrak{D}\right) }\equiv \pm \left[ \lambda
\left( c_{V}-c_{S}\right) -\frac{i}{2}\left( c_{V}^{2}-c_{S}^{2}\right)
\left( S_{-}\pm D_{-}\right) \right] \;.%
\end{array}
\label{PpmSD}
\end{equation}

The transmission and reflection coefficients are thus obtained by solving
the system (\ref{Eqn_arr3})%
\begin{equation}
\begin{array}{c}
T_{R\rightarrow L}=\frac{\mathfrak{S-D}}{2\widetilde{h}\left( k-b\right) }=%
\frac{\det M_{-}\left( 2\det M_{-}+i\widetilde{g}\left( a-k\right) 
\widetilde{h}\left( k+b\right) \left( \mathfrak{P}_{+}^{\left( \mathfrak{D}%
\right) }+\mathfrak{P}_{+}^{\left( \mathfrak{S}\right) }\right) \right) }{d_{%
\mathfrak{S}}}\;, \\ 
R_{R\rightarrow L}=\frac{\mathfrak{S+D}}{2\widetilde{h}\left( k+b\right) }=%
\frac{i\widetilde{g}\left( a-k\right) \widetilde{h}\left( k-b\right) \det
M_{-}\left( \mathfrak{P}_{+}^{\left( \mathfrak{D}\right) }-\mathfrak{P}%
_{+}^{\left( \mathfrak{S}\right) }\right) }{d_{\mathfrak{S}}}\;,%
\end{array}
\label{TR_RL}
\end{equation}%
with%
\begin{equation}
\begin{array}{c}
d_{\mathfrak{S}}=2\left( \det M_{-}\right) ^{2}+i\widetilde{g}\left(
a-k\right) \widetilde{h}\left( k+b\right) \det M_{-}\left( \mathfrak{P}%
_{+}^{\left( \mathfrak{D}\right) }+\mathfrak{P}_{+}^{\left( \mathfrak{S}%
\right) }\right) \\ 
-i\widetilde{g}\left( a+k\right) \widetilde{h}\left( k-b\right) \det
M_{-}\left( \mathfrak{P}_{-}^{\left( \mathfrak{D}\right) }-\mathfrak{P}%
_{-}^{\left( \mathfrak{S}\right) }\right) \\ 
+\widetilde{g}\left( a+k\right) \widetilde{g}\left( a-k\right) \widetilde{h}%
\left( k+b\right) \widetilde{h}\left( k-b\right) \left( \mathfrak{P}%
_{+}^{\left( \mathfrak{S}\right) }\mathfrak{P}_{-}^{\left( \mathfrak{D}%
\right) }-\mathfrak{P}_{-}^{\left( \mathfrak{S}\right) }\mathfrak{P}%
_{+}^{\left( \mathfrak{D}\right) }\right)%
\end{array}
\label{d_Sigma}
\end{equation}

\bigskip Formulae (\ref{TR_RL}) can be further semplified by noting that%
\begin{equation*}
\begin{array}{c}
\mathfrak{P}_{+}^{\left( \mathfrak{D}\right) }+\mathfrak{P}_{+}^{\left( 
\mathfrak{S}\right) }=2\frac{c_{V}E+c_{S}m}{k}-2i\left(
c_{V}^{2}-c_{S}^{2}\right) N_{-}^{\left( 1\right) }\;, \\ 
\mathfrak{P}_{-}^{\left( \mathfrak{D}\right) }-\mathfrak{P}_{-}^{\left( 
\mathfrak{S}\right) }=-2\frac{c_{V}E+c_{S}m}{k}+2i\left(
c_{V}^{2}-c_{S}^{2}\right) N_{-}^{\left( 2\right) }\;, \\ 
\mathfrak{P}_{+}^{\left( \mathfrak{S}\right) }\mathfrak{P}_{-}^{\left( 
\mathfrak{D}\right) }-\mathfrak{P}_{-}^{\left( \mathfrak{S}\right) }%
\mathfrak{P}_{+}^{\left( \mathfrak{D}\right) }=-2\left(
c_{V}^{2}-c_{S}^{2}\right) \det M_{-}\;.%
\end{array}%
\end{equation*}

It turns out that%
\begin{equation}
\begin{array}{c}
\frac{d_{\mathfrak{S}}}{\det M_{-}}=2\left\{ \det M_{-}+i\frac{c_{V}E+c_{S}m%
}{k}\left[ \widetilde{g}\left( a-k\right) \widetilde{h}\left( k+b\right) +%
\widetilde{g}\left( a+k\right) \widetilde{h}\left( k-b\right) \right] \right.
\\ 
\left. +\left( c_{V}^{2}-c_{S}^{2}\right) \left[ \widetilde{g}\left(
a-k\right) \widetilde{h}\left( k+b\right) N_{-}^{\left( 1\right) }+%
\widetilde{g}\left( a+k\right) \widetilde{h}\left( k-b\right) N_{-}^{\left(
2\right) }\right. \right. \\ 
\left. \left. -\widetilde{g}\left( a+k\right) \widetilde{g}\left( a-k\right) 
\widetilde{h}\left( k+b\right) \widetilde{h}\left( k-b\right) \right]
\right\} =2\det M_{+}\;.%
\end{array}
\label{detM_pm_rel}
\end{equation}

The last step is proved in detail in Appendix B. With the above result,
formulae (\ref{TR_RL}) are written as%
\begin{eqnarray}
T_{R\rightarrow L} &=&\frac{\det M_{-}+\widetilde{g}\left( a-k\right) 
\widetilde{h}\left( k+b\right) \left[ i\frac{c_{V}E+c_{S}m}{k}+\left(
c_{V}^{2}-c_{S}^{2}\right) N_{-}^{\left( 1\right) }\right] }{\det M_{+}}\;,
\label{TR_RL_fin} \\
R_{R\rightarrow L} &=&-\frac{\widetilde{g}\left( a-k\right) \widetilde{h}%
\left( k-b\right) \left[ i\frac{c_{V}E+c_{S}m}{k}+\left(
c_{V}^{2}-c_{S}^{2}\right) N_{-}^{\left( 2\right) }\right] }{\det M_{+}}\;. 
\notag
\end{eqnarray}

It is straightforward to verify that\cite{CV06} 
\begin{equation}
\begin{array}{c}
T_{L\rightarrow R}\left( -a,-b\right) =T_{R\rightarrow L}\left( a,b\right)
\;, \\ 
R_{L\rightarrow R}\left( -a,-b\right) =R_{R\rightarrow L}\left( a,b\right)
\;.%
\end{array}%
\end{equation}%
The scattering matrix, $S$, can be defined as in Ref.\cite{CDV07}%
\begin{equation}
S=\left( 
\begin{array}{cc}
T_{L\rightarrow R} & R_{R\rightarrow L} \\ 
R_{L\rightarrow R} & T_{R\rightarrow L}%
\end{array}%
\right) \;.  \label{S_matr}
\end{equation}

The general properties of the $S$ matrix obtained in Ref.\cite{CDV07} in
case of $\mathcal{P}$, $\mathcal{T}$, or $\mathcal{PT}$ invariance of the
Hamiltonian hold in relativistic quantum mechanics, too. In particular, $%
\mathcal{PT}$ symmetry of the Hamiltonian implies%
\begin{equation}
S^{-1}=S^{\ast }\;,
\end{equation}%
or%
\begin{equation}
\begin{array}{c}
|\det S|=1\;, \\ 
|T_{L\rightarrow R}|=|T_{R\rightarrow L}|\;, \\ 
Im(R_{L\rightarrow R}R_{R\rightarrow L}^{\ast })=0\;.%
\end{array}
\label{PT_cond}
\end{equation}

$T_{L\rightarrow R}$ and $T_{R\rightarrow L}$ have the same modulus, but
different phase: the latter property, characteristic of non-local
potentials, is discussed in particular in Refs.\cite{CV06},\cite{CDV07}.

Finally, the last of conditions (\ref{PT_cond}) implies that $%
R_{R\rightarrow L}$ and $R_{L\rightarrow R}$ have the same phase, although
they have different moduli, since unitarity is broken.

\section{Symmetries and non-relativistic limits}

Eq. (\ref{Dirac_stat}) is equivalent to a pair of coupled differential
equations in the two components of the Dirac spinor $\Psi \left( x\right)
=\left( 
\begin{array}{c}
\Psi _{1}\left( x\right) \\ 
\Psi _{2}\left( x\right)%
\end{array}%
\right) $; in the Dirac representation, where $\alpha _{x}=\sigma _{x}$ and $%
\beta =\sigma _{z}$%
\begin{equation}
\left\{ 
\begin{array}{c}
\left( m-E\right) \Psi _{1}\left( x\right) -i\frac{\partial }{\partial x}%
\Psi _{2}\left( x\right) +\left( c_{S}+c_{V}\right) \int_{-\infty }^{+\infty
}dyK\left( x,y\right) \Psi _{1}\left( y\right) =0 \\ 
-i\frac{\partial }{\partial x}\Psi _{1}\left( x\right) -\left( m+E\right)
\Psi _{2}\left( x\right) +\left( c_{V}-c_{S}\right) \int_{-\infty }^{+\infty
}dyK\left( x,y\right) \Psi _{2}\left( y\right) =0%
\end{array}%
\right. \;.  \label{Dirac_sys}
\end{equation}

For arbitrary values of the coupling strengths, $c_{S}$ and $c_{V}$, the
above equations do not decouple; decoupling occurs when $c_{V}=\pm c_{S}$. \
The method of solution described in the preceding section remains valid and
the final results for the reflection and transmission coefficients are still
given by formulae (\ref{T_LR}-\ref{R_LR}) for progressive waves and by
formulae (\ref{TR_RL}) for regressive waves, even if intermediate formulae
are different.

In $3+1$ dimensions, the cases $c_{V}=c_{S}$ and $c_{V}=-c_{S}$ are examples
of Bell-Ruegg symmetries\cite{BR75}, where the Dirac Hamiltonian commutes
with the generators of an $SU(2)$ group, constructed with Dirac matrices and
the momentum operator. The eigenstates of the Dirac Hamiltonian belong to
the carrier space of the spinor representation of such a group and are thus
doubly degenerate. When $c_{V}=c_{S}$, \ the members of the doublet have the
same radial quantum number $n_{r}$, the same orbital momentum $l$ and total
angular momentum $j=l\pm \frac{1}{2}$ (spin symmetry). When $c_{V}=-c_{S}$,
they have quantum numbers $\left( n_{r},l,j=l+\frac{1}{2}\right) $ and $%
\left( n_{r}-1,l+2,j=l+\frac{3}{2}\right) $, \textit{i.e. }the\textit{\ }%
same pseudo-orbital momentum $\widetilde{l}=l+1$ and pseudo-spin $\widetilde{%
s}=\frac{1}{2}$, so that $j=\widetilde{l}\pm \frac{1}{2}$ (pseudo-spin
symmetry). The mean field of heavy nuclei exhibits an approximate
pseudo-spin symmetry, experimentally known for many years, but correctly
explained as a relativistic effect only few years ago\cite{Gi97}. At a
phenomenological level, the approximate pseudo-spin symmetry naturally
arises in relativistic mean field models, where the nuclear mean field is in
practice the sum of an attractive scalar field (the $\sigma $ field) and of
a repulsive vector field (the $\omega $ field) of almost the same strength.
At a more fundamental level, it can be obtained from sum rules of quantum
chromodynamics in nuclear matter\cite{Gi05}.

Let us consider the case $c_{V}=c_{S}=c$ first. We promptly obtain in this
case 
\begin{equation}
\left\{ 
\begin{array}{c}
-\frac{\partial ^{2}}{\partial x^{2}}\Psi _{1}\left( x\right) +2c\left(
m+E\right) \int_{-\infty }^{+\infty }dyK\left( x,y\right) \Psi _{1}\left(
y\right) =\left( E^{2}-m^{2}\right) \Psi _{1}\left( x\right) \equiv
k^{2}\Psi _{1}\left( x\right) \\ 
\Psi _{2}\left( x\right) =\frac{-i}{m+E}\frac{\partial }{\partial x}\Psi
_{1}\left( x\right)%
\end{array}%
\right. \;.  \label{c_v=c_s}
\end{equation}

The above system is suited to the study of the non-relativistic limit ($%
E\rightarrow m+\frac{k^{2}}{2m}$ , with $\frac{k^{2}}{2m}\ll m$), where the
first equation of system (\ref{c_v=c_s}), satisfied by $\Psi _{1}$, becomes
a Schr\"{o}dinger equation with a non-local potential of strength $s=2c$ and
kernel $K$. $\Psi _{2}$, being proportional to $\frac{\partial }{\partial x}%
\Psi _{1}$, does not obey a Schr\"{o}dinger-like equation.

In this limit, the transmission and reflection coefficients obtained in the
preceding section simplify considerably. In fact, from formulae (\ref{T_LR}-%
\ref{R_LR}) we promptly obtain, for $c_{V}=c_{S}$ $=c$ and $E\rightarrow m+%
\frac{k^{2}}{2m}$%
\begin{equation}
\left\{ 
\begin{array}{c}
\lim_{E\rightarrow m+\frac{k^{2}}{2m}}T_{L\rightarrow R}=1-i\frac{2cm}{k}%
\frac{\widetilde{g}\left( k-a\right) \widetilde{h}\left( k+b\right) }{1+i%
\frac{2cm}{k}S_{+}} \\ 
\lim_{E\rightarrow m+\frac{k^{2}}{2m}}R_{L\rightarrow R}=-i\frac{2cm}{k}%
\frac{\widetilde{g}\left( k+a\right) \widetilde{h}\left( k+b\right) }{1+i%
\frac{2cm}{k}S_{+}}%
\end{array}%
\right. \;,
\end{equation}%
in agreement with formulae (153) of Ref.\cite{CDV07}, where $\frac{2cm}{k}$
is indicated with $\omega $ and $\frac{1}{1+i\frac{2cm}{k}S_{+}}$ with $%
D_{+} $, not to be confused with the $D_{+}$ integral defined in formulae (%
\ref{S_D}) of the preceding section. It is worthwhile to recall that Ref.%
\cite{CDV07} uses units $2m=1$, as is common in non-relativistic quantum
mechanics.

In the same way, we obtain, after some simple algebra%
\begin{equation}
\left\{ 
\begin{array}{c}
\lim_{E\rightarrow m+\frac{k^{2}}{2m}}T_{R\rightarrow L}=1-i\frac{2cm}{k}%
\frac{\widetilde{g}\left( k+a\right) \widetilde{h}\left( k-b\right) }{1+i%
\frac{2cm}{k}\left[ -S_{-}+\widetilde{g}\left( k-a\right) \widetilde{h}%
\left( k+b\right) +\widetilde{g}\left( k+a\right) \widetilde{h}\left(
k-b\right) \right] } \\ 
\lim_{E\rightarrow m+\frac{k^{2}}{2m}}R_{R\rightarrow L}=-i\frac{2cm}{k}%
\frac{\widetilde{g}\left( k-a\right) \widetilde{h}\left( k-b\right) }{1+i%
\frac{2cm}{k}\left[ -S_{-}+\widetilde{g}\left( k-a\right) \widetilde{h}%
\left( k+b\right) +\widetilde{g}\left( k+a\right) \widetilde{h}\left(
k-b\right) \right] }%
\end{array}%
\right. \;,
\end{equation}%
which coincide with formulae (156) of Ref.\cite{CDV07}, where $i\frac{%
2cmS_{-}}{k}$ is indicated with $N_{-}$, not to be confused with the $N_{-}$
matrix defined in formula (\ref{Npm_ter}).

In the case $c_{V}=-c_{S}=c^{\prime }$ $\Psi _{1}$and $\Psi _{2}$
interchange their role, since the two decoupled equations now are%
\begin{equation}
\left\{ 
\begin{array}{c}
\Psi _{1}\left( x\right) =\frac{-i}{E-m}\frac{\partial }{\partial x}\Psi
_{2}\left( x\right) \\ 
-\frac{\partial ^{2}}{\partial x^{2}}\Psi _{2}\left( x\right) +2c^{\prime
}\left( E-m\right) \int_{-\infty }^{+\infty }dyK\left( x,y\right) \Psi
_{2}\left( y\right) =\left( E^{2}-m^{2}\right) \Psi _{2}\left( x\right)
\equiv k^{2}\Psi _{2}\left( x\right)%
\end{array}%
\right. \;.  \label{c_V=-c_S}
\end{equation}

The formulae of transmission and reflection coefficients now depend on $E-m$%
, to be replaced in the non-relativistic limit by the kinetic energy $\frac{%
k^{2}}{2m}$. In that limit, the second equation (\ref{c_V=-c_S}), satisfied
by $\Psi _{2}$, becomes a Schr\H{o}dinger equation with an energy dependent
coupling strength $s\left( k\right) =c^{\prime }k^{2}/(2m^{2})$, while $\Psi
_{1}$ is not solution to a Schr\H{o}dinger equation.

The final expressions are%
\begin{equation}
\left\{ 
\begin{array}{c}
\lim_{E\rightarrow m+\frac{k^{2}}{2m}}T_{L\rightarrow R}=1-\frac{ic^{\prime
}k}{2m}\frac{\widetilde{g}\left( k-a\right) \widetilde{h}\left( k+b\right) }{%
1+i\frac{c^{\prime }k}{2m}S_{+}} \\ 
\lim_{E\rightarrow m+\frac{k^{2}}{2m}}R_{L\rightarrow R}=i\frac{c^{\prime }k%
}{2m}\frac{\widetilde{g}\left( k+a\right) \widetilde{h}\left( k+b\right) }{%
1+i\frac{c^{\prime }k}{2m}S_{+}}%
\end{array}%
\right.
\end{equation}%
and%
\begin{equation}
\left\{ 
\begin{array}{c}
\lim_{E\rightarrow m+\frac{k^{2}}{2m}}T_{R\rightarrow L}=1-i\frac{c^{\prime
}k}{2m}\frac{\widetilde{g}\left( k+a\right) \widetilde{h}\left( k-b\right) }{%
1+\frac{ic^{\prime }k}{2m}\left[ -S_{-}+\widetilde{g}\left( k-a\right) 
\widetilde{h}\left( k+b\right) +\widetilde{g}\left( k+a\right) \widetilde{h}%
\left( k-b\right) \right] } \\ 
\lim_{E\rightarrow m+\frac{k^{2}}{2m}}R_{R\rightarrow L}=i\frac{c^{\prime }k%
}{2m}\frac{\widetilde{g}\left( k-a\right) \widetilde{h}\left( k-b\right) }{1+%
\frac{ic^{\prime }k}{2m}\left[ -S_{-}+\widetilde{g}\left( k-a\right) 
\widetilde{h}\left( k+b\right) +\widetilde{g}\left( k+a\right) \widetilde{h}%
\left( k-b\right) \right] }%
\end{array}%
\right. \;.
\end{equation}

As expected, the above formulae have the same structure as \ those in the
case $c_{V}=c_{S}$, with the constant strength $s=2c$ replaced with the
energy-dependent strength $s\left( k\right) =c^{\prime }k^{2}/(2m^{2})$.

For arbitrary values of $c_{V}$ and $c_{S}$ the equations (\ref{Dirac_sys})
do not decouple, unless the potential becomes local, $K(x,y)=\delta
(x-y)V(x) $. As a consequence, in the particular case of a purely scalar
potential, $c_{V}=0$, we do not obtain the pseudo-supersymmetric scheme of
Ref.\cite{SR05}, which holds for local potentials only.

Summing up, the cases $c_{V}=\pm c_{S}$, reflecting the Bell-Ruegg symmetries%
\cite{BR75} in one dimension, reduce the two-dimensional manifold $[\Psi
_{1},\Psi _{2}]$ to the one-dimensional manifold $\Psi _{1}$ when $%
c_{V}=c_{S}$, or $\Psi _{2}$ when $c_{V}=-c_{S}$, the latter case being
relevant for nuclear physics.

\section{Bound states with real energy}

The $\mathcal{PT}$ symmetry of the potential kernel, $K$, permits the
general statement that either bound state energies are real, or that they
come in complex conjugate pairs: in fact, if $\Psi \left( x\right) $ is the
solution to the stationary Dirac equation with energy $E$ and momentum $%
p_{x} $ 
\begin{equation}
(E-\alpha _{x}p_{x}-\beta m)\Psi \left( x\right) -\left( c_{S}\beta
+c_{V}\right) \int\limits_{-\infty }^{+\infty }dyK(x,y)\Psi \left( y\right)
=0\;,
\end{equation}%
$\mathcal{PT}\Psi \left( x\right) =\Psi ^{\ast }\left( -x\right) $ is
solution to the Dirac equation with energy $E^{\ast }$ and momentum $%
-p_{x}^{\ast }$%
\begin{equation}
(E^{\ast }+\alpha _{x}p_{x}^{\ast }-\beta m)\mathcal{PT}\Psi \left( x\right)
-\left( c_{S}\beta +c_{V}\right) \int\limits_{-\infty }^{+\infty }dyK^{\ast
}(-x,-y)\mathcal{PT}\Psi \left( y\right) =0\;.
\end{equation}

We now treat in particular bound states $\Psi _{bs}\left( x\right) $ with
real energy and imaginary momentum $p_{x}=-p_{x}^{\ast }$ and investigate
the relation between $\Psi _{bs}\left( x\right) $ and $\mathcal{PT}\Psi
_{bs}\left( x\right) $. From now on, the quantum number $k$ is no more real
and positive, as defined in Section 2, but complex.

The Green function formalism permits not only derivation of scattering, but
also of bound state wave functions. As is known, bound state energies are
located in the interval $-m<E<+m$, where the square of the momentum, $%
k^{2}=E^{2}-m^{2}$, is negative, \textit{i. e., }$k=i\overline{k}$ is
imaginary. Bound state wave functions can be obtained by analytic
continuation of one of the two independent scattering solutions, \textit{e.
g. }$\Psi _{+}\left( x\right) $ from formula (\ref{Psi_pm}), to the positive
imaginary $k$ axis, \textit{i. e. }we can take $\overline{k}=\sqrt{%
m^{2}-E^{2}}>0$, and impose the boundary conditions $\lim_{x\rightarrow \pm
\infty }\Psi _{+}\left( x\right) =0$. Owing to the fact that the Green
function $G_{+}$ vanishes at $x=\pm \infty $ when $k=i\overline{k}$, we must
get rid of the free-wave contribution, by putting $A_{+}=B_{+}=0$. We thus
obtain%
\begin{equation}
\Psi _{bs}\left( x\right) =-\int\limits_{-\infty }^{+\infty }dx^{\prime
}g\left( x^{\prime }\right) e^{iax^{\prime }}G_{+}\left( x-x^{\prime
}\right) \left( c_{S}\sigma _{z}+c_{V}\right) I_{+}.\;\;\;\left( k=i%
\overline{k}\right) \;  \label{Psi_bs}
\end{equation}

Remembering expression (\ref{Gpm}) for $G_{+}\left( x-x^{\prime }\right) $ , 
$\Psi _{bs}\left( x\right) $ can be put in the form%
\begin{equation}
\begin{array}{c}
\Psi _{bs}\left( x\right) =-\frac{1}{2\overline{k}}\left\{ e^{-\overline{k}x}%
\mathcal{I}_{1}\left( x\right) \left( i\overline{k}\sigma _{x}+m\sigma
_{z}+E\right) \right. \\ 
\left. +e^{\overline{k}x}\mathcal{I}_{2}\left( x\right) \left( -i\overline{k}%
\sigma _{x}+m\sigma _{z}+E\right) \right\} \left( c_{S}\sigma
_{z}+c_{V}\right) I_{+}\;,%
\end{array}%
\end{equation}%
where%
\begin{equation}
\mathcal{I}_{1}\left( x\right) \equiv \int_{-\infty }^{x}dx^{\prime }g\left(
x^{\prime }\right) e^{\left( ia+\overline{k}\right) x^{\prime }}  \label{i_1}
\end{equation}%
and%
\begin{equation}
\mathcal{I}_{2}\left( x\right) \equiv \int_{x}^{+\infty }dx^{\prime }g\left(
x^{\prime }\right) e^{\left( ia-\overline{k}\right) x^{\prime }}\;.
\label{i_2}
\end{equation}%
The normalizability of $\Psi _{bs}$ is easily cheched by noting that $%
\lim_{x\rightarrow \pm \infty }e^{-\overline{k}x}\mathcal{I}%
_{1}(x)=\lim_{x\rightarrow \pm \infty }e^{\overline{k}x}\mathcal{I}_{2}(x)=0$%
.

From definitions (\ref{i_1}-\ref{i_2}), remembering that $g\left( x^{\prime
}\right) =g\left( -x^{\prime }\right) $, it is easy to verify that 
\begin{equation}
\mathcal{I}_{2}\left( x\right) =\mathcal{I}_{1}^{\ast }\left( -x\right) =%
\mathcal{PT}\mathcal{I}_{1}\left( x\right) \;.  \label{PTi_1}
\end{equation}

The integral equation (\ref{Psi_bs}) allows us to compute bound state
energies, too. By multiplying both sides by $h\left( x\right) e^{ibx}$ and
integrating them over $x$ from $-\infty $ to $+\infty $, we obtain%
\begin{equation}
I_{+}=-\int_{-\infty }^{+\infty }dxh\left( x\right) e^{ibx}\int_{-\infty
}^{+\infty }dx^{\prime }g\left( x^{\prime }\right) e^{iax^{\prime
}}G_{+}\left( x-x^{\prime }\right) \left( c_{S}\sigma _{z}+c_{V}\right)
I_{+}\;,
\end{equation}%
or, remembering definition (\ref{Mpm}) of matrix $M_{+}$ 
\begin{eqnarray}
&&(1+\int_{-\infty }^{+\infty }dxh(x)e^{ibx}\int_{-\infty }^{+\infty
}dx^{\prime }g(x^{\prime })e^{iax^{\prime }}G_{+}(x-x^{\prime })(c_{S}\sigma
_{z}+c_{V}))I_{+}  \notag \\
&\equiv &\left( 
\begin{array}{cc}
M_{+}^{11} & M_{+}^{12} \\ 
M_{+}^{21} & M_{+}^{22}%
\end{array}%
\right) \left( 
\begin{array}{c}
I_{+}^{1} \\ 
I_{+}^{2}%
\end{array}%
\right) =0\;.  \label{IM_0}
\end{eqnarray}

Note that, since $M_{\pm }\left( i\overline{k}\right) =\mathcal{PT}M_{\pm
}\left( i\overline{k}\right) \left( \mathcal{PT}\right) ^{-1}$, if $I_{\pm }$
is solution of Eq. (\ref{IM_0}), $\mathcal{PT}I_{\pm }$ is solution, too.

The necessary condition for a non-trivial solution of the above equation

\begin{equation}
\det M_{+}=1+\frac{c_{V}E+c_{S}m}{\sqrt{m^{2}-E^{2}}}S_{+}+\frac{\left(
c_{V}^{2}-c_{S}^{2}\right) }{4}\left( D_{+}^{2}-S_{+}^{2}\right) =0\;\;,
\label{detM_zero}
\end{equation}%
where $S_{+}$ and $D_{+}$ are functions of $k(E)$, fixes bound state
energies as the roots of the equation in the interval $-m<E<+m$. Not
surprisingly, bound states correspond to poles of the transmission
coefficient $T_{L\rightarrow R}$ (\ref{T_LR}).

Eq. (\ref{IM_0}) allows one to express the ratio of the components of $I_{+}$
$\ $in terms of $M_{+}$ matrix elements. In general, one observes that $%
M_{+}^{22}=1+i\frac{\lambda }{2}\left( c_{V}-c_{S}\right) S_{+}\neq 0$, so
that one can exploit the second eq. (\ref{IM_0}), which gives $%
I_{+}^{2}=-\left( M_{+}^{21}/M_{+}^{22}\right) I_{+}^{1}$ and $\Psi
_{bs}\left( x\right) $ can be written as%
\begin{equation}
\begin{array}{c}
\Psi _{bs}\left( x\right) =-\frac{I_{+}^{1}}{2\overline{k}}\left\{ e^{-%
\overline{k}x}\mathcal{I}_{1}\left( x\right) \left( 
\begin{array}{cc}
E+m & i\overline{k} \\ 
i\overline{k} & E-m%
\end{array}%
\right) \right. \\ 
\left. +e^{\overline{k}x}\mathcal{I}_{2}\left( x\right) \left( 
\begin{array}{cc}
E+m & -i\overline{k} \\ 
-i\overline{k} & E-m%
\end{array}%
\right) \right\} \left( 
\begin{array}{cc}
c_{V}+c_{S} & 0 \\ 
0 & c_{V}-c_{S}%
\end{array}%
\right) \left( 
\begin{array}{c}
1 \\ 
-\frac{i}{2}\frac{\left( c_{V}-c_{S}\right) D_{+}}{1+i\frac{\lambda }{2}%
\left( c_{V}-c_{S}\right) S_{+}}%
\end{array}%
\right) \;.%
\end{array}
\label{Psi_bs_gen}
\end{equation}

In the above formula, the modulus of $I_{+}^{1}$ is to be determined from
normalization of the wave function $\Psi _{bs}\left( x\right) $. It is easy
to check that $\Psi _{bs}\left( x\right) $ is eigenstate of $\mathcal{PT}$,
since the matrices in curly brackets are $\mathcal{PT}$-symmetric, owing to
relation (\ref{PTi_1}) and the ratio $r=-\frac{i}{2}\frac{\left(
c_{V}-c_{S}\right) D_{+}}{1+i\frac{\lambda }{2}\left( c_{V}-c_{S}\right)
S_{+}}$ is real. In fact, from definitions (\ref{N1pm}-\ref{N2pm}), we see
that, for $k=i\overline{k}$, $N_{+}^{(2)}\left( i\overline{k}\right) =\left(
N_{+}^{(1)}\left( i\overline{k}\right) \right) ^{\ast }$; thus, $S_{+}\left(
i\overline{k}\right) \equiv N_{+}^{(1)}\left( i\overline{k}\right)
+N_{+}^{(2)}\left( i\overline{k}\right) =2Re\left( N_{+}^{(1)}\left( i%
\overline{k}\right) \right) $ is real, $D_{+}\left( i\overline{k}\right)
\equiv N_{+}^{(1)}\left( i\overline{k}\right) -N_{+}^{(2)}\left( i\overline{k%
}\right) =2i Im\left( N_{+}^{(1)}\left( i\overline{k}\right) \right) $ is
imaginary and $\lambda =i\overline{k}/\left( E+m\right) $ is imaginary, too,
so that, as a final result, $r$ is real and $\mathcal{PT}\Psi _{bs}\left(
x\right) =\Psi _{bs}\left( x\right) $, if $I_{+}^{1}$ is chosen to be real.

Considering that $D_{+}$ and $S_{+}$ do not depend on $c_{V}$, or $c_{S}$,
Eq. (\ref{detM_zero}) can also be used to determine either potential
strength ($c_{V}$ or $c_{S}$), provided the other is fixed, in particular
set to zero, in such a way to obtain a bound state at a given energy $E$ in
the $(-m,+m)$ range. In this procedure, however, $\mathcal{PT}$ symmetry is
not automatically preserved, since Eq. (\ref{detM_zero}) is of second degree
in the unknown potential strength and might have a pair of complex conjugate
solutions.

We have derived our expressions for bound-state wave functions starting from 
$\Psi _{+}\left( x\right) $, but we could, alternatively, start from $\Psi
_{-}\left( x\right) $ and determine the constants $A_{-}$ and $B_{-}$ from
the boundary conditions $\lim_{x\rightarrow \pm \infty }\Psi _{-}\left(
x\right) =0$. In this case, both $A_{-}$ and $B_{-}$ must be different from
zero because of the asymptotic behaviour of $G_{-}$ and appear as the
solution of a system of two homogeneous linear equations. The condition for
a non-trivial solution of the system yields again the equation $\det
M_{+}\left( i\overline{k}\right) =0$, as expected, with $\det M_{+}$ written
in terms of $\det M_{-}$ according to formula (\ref{detM_pm_rel}).

\section{The Yamaguchi potential}

As an example of application of the formalism developed in the preceding
sections, we now work out in detail a one-dimensional $\mathcal{PT}$%
-symmetric version of the Yamaguchi potential\cite{Ya54}, originally aimed
at describing bound and scattering states of the neutron-proton system. \ We
assume 
\begin{equation}
g\left( x\right) =\exp \left( -c\left\vert x\right\vert \right) ,\;h\left(
y\right) =\exp \left( -d\left\vert y\right\vert \right) ,\;\left( -\infty
<x,y<+\infty \right)
\end{equation}%
with $c$ and $d$ positive constants, so that the Fourier transforms are%
\begin{equation}
\widetilde{g}\left( q\right) =\frac{2c}{c^{2}+q^{2}},\;\widetilde{h}\left(
q^{\prime }\right) =\frac{2d}{d^{2}+q^{\prime 2}}.\;\left( -\infty
<q,q^{\prime }<+\infty \right)  \label{g_h_tilde}
\end{equation}%
and the $\mathcal{PT}$-symmetric kernel reads%
\begin{equation}
K\left( x,y\right) =e^{-c\left\vert x\right\vert +iax}e^{-d\left\vert
y\right\vert +iby},
\end{equation}%
with $a$ and $b$ real constants. With the above definitions the basic
integrals $N_{\pm }^{\left( 1\right) }$ from formula (\ref{N1pm}) and $%
N_{\pm }^{\left( 2\right) }$ from formula (\ref{N2pm}), as well as their
linear combinations $S_{\pm }=N_{\pm }^{\left( 1\right) }+N_{\pm }^{\left(
2\right) }$ and $D_{\pm }=N_{\pm }^{\left( 1\right) }-N_{\pm }^{\left(
2\right) }$, can be computed by elementary methods. We only quote the final
results%
\begin{equation}
N_{\pm }^{\left( 1\right) }=-\frac{i}{c}\widetilde{g}\left( a\mp k\right) 
\frac{\left( a+b\right) c+\left( c+d\right) \left( a\mp k\right) }{\left(
a+b\right) ^{2}+\left( c+d\right) ^{2}}+\frac{\widetilde{g}\left( a\mp
k\right) \widetilde{h}\left( b\pm k\right) }{2}\left( 1+i\frac{b\pm k}{d}%
\right) \;,
\end{equation}%
\begin{equation}
N_{\pm }^{\left( 2\right) }=\frac{i}{c}\widetilde{g}\left( a\pm k\right) 
\frac{\left( a+b\right) c+\left( c+d\right) \left( a\pm k\right) }{\left(
a+b\right) ^{2}+\left( c+d\right) ^{2}}+\frac{\widetilde{g}\left( a\pm
k\right) \widetilde{h}\left( b\mp k\right) }{2}\left( 1-i\frac{b\mp k}{d}%
\right) \;.  \label{N12_Ya}
\end{equation}

Formulae (\ref{N12_Ya}) become particularly simple when applied to the
analysis of bound states: in this case, we already know from the previous
section that $N_{\pm }^{\left( 2\right) }\left( i\overline{k}\right) =\left(
N_{\pm }^{\left( 1\right) }\left( i\overline{k}\right) \right) ^{\ast }$.
Therefore, $S_{+}\equiv N_{+}^{\left( 1\right) }+N_{+}^{\left( 2\right)
}=2Re\left( N_{+}^{\left( 1\right) }\right) $ and $D_{+}\equiv N_{+}^{\left(
1\right) }-N_{+}^{\left( 2\right) }=2iIm\left( N_{+}^{\left( 1\right)
}\right) $; the left-hand-side of Eq. (\ref{detM_zero}) thus becomes real in
the interval $-m<E<+m$: bound state energies are roots of the real equation%
\begin{equation}
\det M_{+}=1+\frac{2\left( c_{V}E+c_{S}m\right) }{\sqrt{m^{2}-E^{2}}}%
Re\left( N_{+}^{\left( 1\right) }\right) -\left( c_{V}^{2}-c_{S}^{2}\right)
\left\vert N_{+}^{\left( 1\right) }\right\vert ^{2}=0  \label{detM_1}
\end{equation}

If we put $c_{S}=0$ in the above equation, this allows us to derive the
strength $c_{V}$ at which the purely vector potential has a bound state at
given real energy $E$; in fact, Eq. (\ref{detM_1}) can be considered as a
quadratic equation in $c_{V}$, with real solutions%
\begin{equation}
c_{V}=\frac{\frac{ERe\left( N_{+}^{\left( 1\right) }\right) }{\sqrt{%
m^{2}-E^{2}}}\pm \sqrt{\frac{E^{2}\left( Re\left( N_{+}^{\left( 1\right)
}\right) \right) ^{2}}{m^{2}-E^{2}}+\left\vert N_{+}^{\left( 1\right)
}\right\vert ^{2}}}{\left\vert N_{+}^{\left( 1\right) }\right\vert ^{2}}\;.
\label{c_v_solv}
\end{equation}

Note that one of the two solutions for $c_{V}$ is always positive.

In order to complete the discussion of bound state wave functions, we give
the corresponding expressions of integrals (\ref{i_1}) and \ref{i_2}),
obtained by elementary integration:%
\begin{equation}
\begin{array}{c}
e^{-\overline{k}x}\mathcal{I}_{1}\left( x\right) =\theta \left( -x\right) 
\frac{e^{\left( c+ia\right) x}}{c+\overline{k}+ia}+\theta \left( x\right) %
\left[ \frac{e^{-\overline{k}x}}{c+\overline{k}+ia}+\frac{e^{\left(
-c+ia\right) x}-e^{-\overline{k}x}}{-c+\overline{k}+ia}\right] \;, \\ 
e^{\overline{k}x}\mathcal{I}_{2}\left( x\right) =\theta \left( -x\right) %
\left[ \frac{e^{\overline{k}x}-e^{\left( c+ia\right) x}}{c-\overline{k}+ia}+%
\frac{e^{\overline{k}x}}{c+\overline{k}-ia}\right] +\theta \left( x\right) 
\frac{e^{\left( -c+ia\right) x}}{c+\overline{k}-ia}\;.%
\end{array}%
\end{equation}

While possible bound state wave functions with real energy are eigenstates
of $\mathcal{PT}$ , scattering wave functions never are, but show some
interesting peculiarities related to transmission resonances when $c_{V}=\pm
c_{S}$, which makes it worthwhile to focus our numerical analysis on that
cases. 
\begin{figure}[h]
\begin{center}
\includegraphics[width=12cm,angle=0,clip]{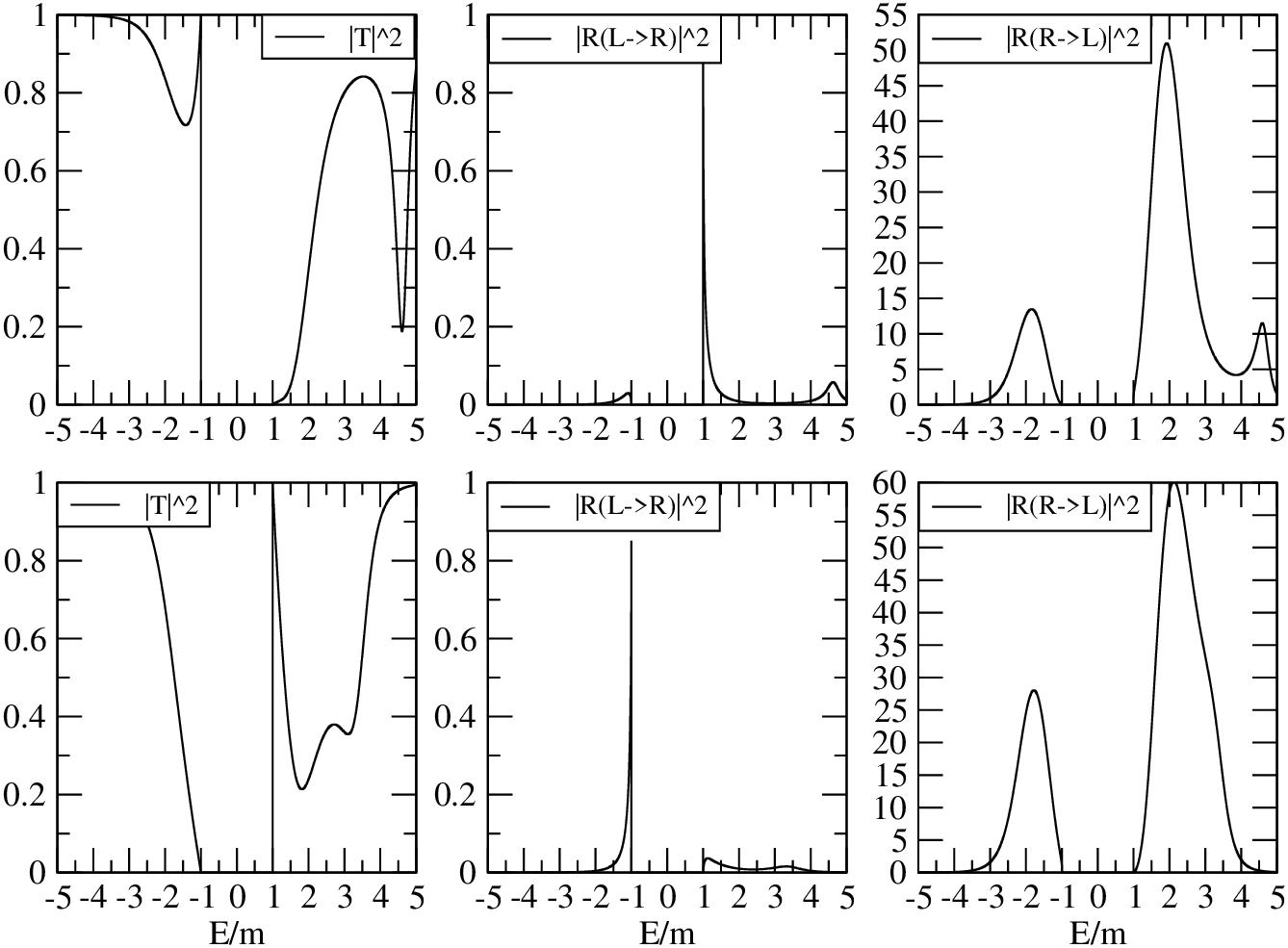}
\end{center}
\caption{Square moduli of transmission and reflection coefficients vs.
energy (in units of particle mass) for Yamaguchi potentials with $a=2$, $b=1$%
, $c=d=1$ and $c_{V}=c_{S}=5m$ (upper panels), or $c_{V}=-c_{S}=5m$ (lower
panels).}
\label{Fig1}
\end{figure}
Figure 1 shows the square moduli of transmission coefficients, $\left\vert
T_{L\rightarrow R}\right\vert ^{2}=\left\vert T_{R\rightarrow L}\right\vert
^{2}\equiv \left\vert T\right\vert ^{2}$, and of reflection coefficients, $%
\left\vert R_{L\rightarrow R}\right\vert ^{2}$ and $\left\vert
R_{R\rightarrow L}\right\vert ^{2}$, as functions of total energy $E$ for
the following choices of potential parameters: $a=2$, $b=1$, $c=d=1$ and $%
c_{V}=c_{S}=5m$ (upper panels), or $c_{V}=-c_{S}=5m$ (lower panels). $E$
ranges from $-5m$ to $+5m$, but the coefficients are not calculated in the $%
-m<E<+m$ interval, where they might have poles corresponding to bound
states. When $c_{V}=c_{S}$, there is a sharp transmission resonance at $E=-m$%
, which appears at $E=+m$ when $c_{V}=-c_{S}$, as expected from the relation 
$T_{L\rightarrow R}\left( c_{V},c_{S},E,k\right) =T_{L\rightarrow R}\left(
-c_{V},c_{S},-E,k\right) $. These zero-energy resonances, called half-bound
states, have $T_{L\rightarrow R}=T_{R\rightarrow L}=1$ and $R_{L\rightarrow
R}=R_{R\rightarrow L}=0$. In both cases, the reflection coefficients show
the handedness discussed in Ref.\cite{Ah04}: the potentials behave as
absorptive for progressive waves ($\left\vert T_{L\rightarrow R}\right\vert
^{2}+\left\vert R_{L\rightarrow R}\right\vert ^{2}<1$) and generative for
regressive waves ( $\left\vert T_{R\rightarrow L}\right\vert ^{2}+\left\vert
R_{R\rightarrow L}\right\vert ^{2}>1$ ). This pattern depends on the
(common) sign of $a$ and $b$ : in fact, owing to the form (\ref{kernel}) of
the kernel, where $g(x)=e^{-c|x|}$ and $h(y)=e^{-d|y|}$ are even functions
of their arguments, changing $a$ into $-a$ and $b$ into $-b$ is equivalent
to a parity transformation ( $x\rightarrow -x$ and $y\rightarrow -y$ ),
namely%
\begin{equation}
\begin{array}{c}
T_{L\rightarrow R}\left( -a,-b\right) =T_{R\rightarrow L}\left( a,b\right)
\;, \\ 
R_{L\rightarrow R}\left( -a,-b\right) =R_{R\rightarrow L}\left( a,b\right)
\;.%
\end{array}%
\end{equation}

In our case, with $a=-2$ and $b=-1$, the potential would become generative
for progressive waves and absorptive for regressive ones. 
\begin{figure}[h]
\begin{center}
\includegraphics[width=12cm,angle=0,clip]{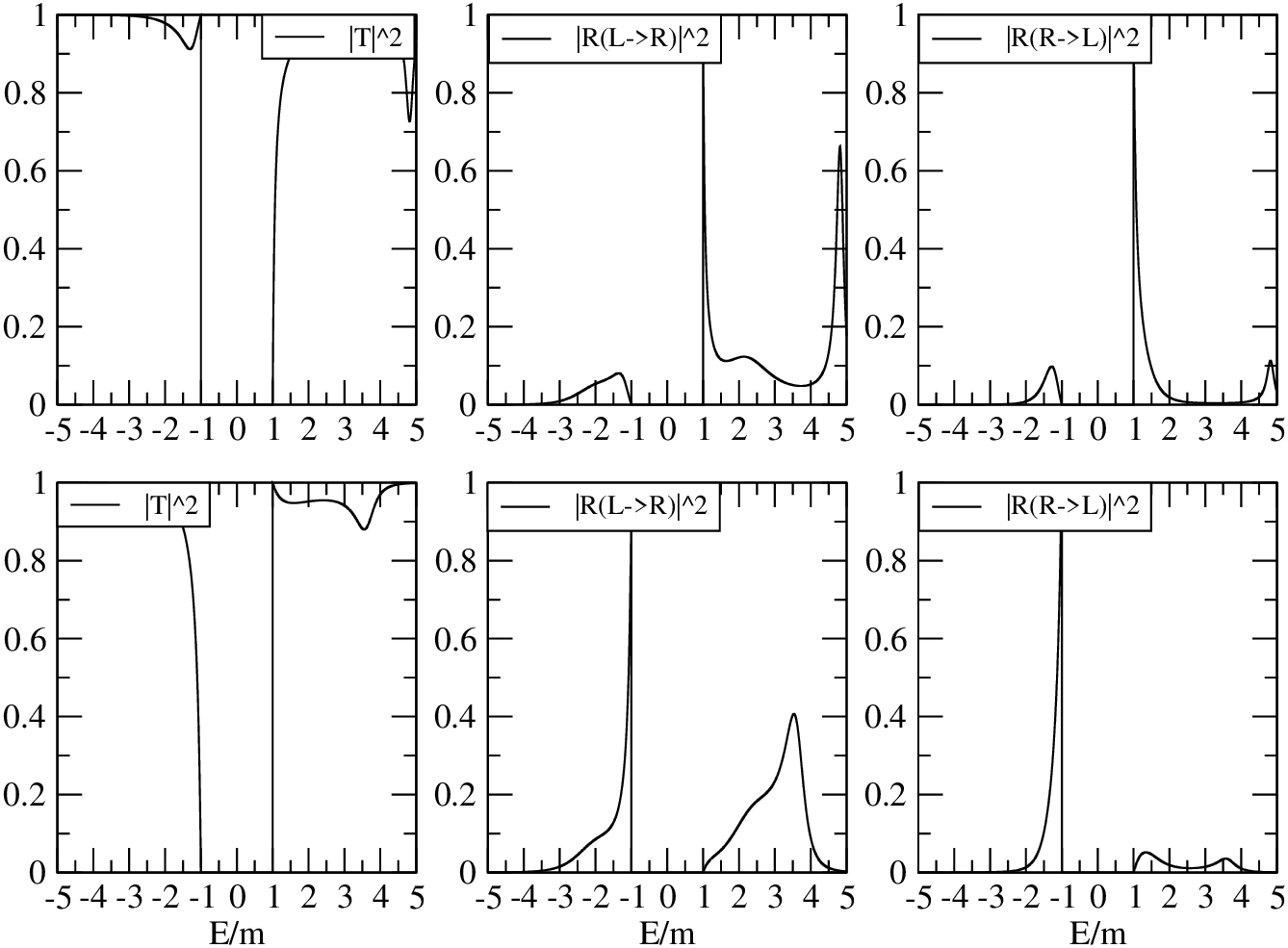}
\end{center}
\caption{Square moduli of transmission and reflection coefficients vs.
energy (in units of particle mass) for Yamaguchi potentials with $a=-2$, $%
b=1 $, $c=d=1$ and $c_V=c_S=2m$ (upper panels), or $c_V=-c_S=2m$ (lower
panels).}
\label{Fig2}
\end{figure}
Handedness, however, is not a general rule: Figure 2 shows transmission and
reflection coefficients for $a=-2$, $b=+1$, $c=d=1$ and $c_{V}=c_{S}=2m$
(upper panels), or $c_{V}=-c_{S}=2m$ (lower panels). In this case, both $%
\left\vert T_{L\rightarrow R}\right\vert ^{2}+\left\vert R_{L\rightarrow
R}\right\vert ^{2}$ and $\left\vert T_{R\rightarrow L}\right\vert
^{2}+\left\vert R_{R\rightarrow L}\right\vert ^{2}$ may be $\lessgtr 1$ in
different energy intervals.

As for bound states, they exist only in the lower panel cases of Figs.1,2:
when $c_{V}=-c_{S}=5m$, $a=2$, $b=1$, $c=d=1$ ( Fig.1 ) there is a real
bound state with energy $\epsilon _{bs}=+0.3835m$, when $c_{V}=-c_{S}=2m$, $%
a=-2$, $b=1$, $c=d=1$ ( Fig.2 ), there is a real bound state at $\epsilon
_{bs}=0.1815m$. If $a$ and $b$ change, the bound states change their
energies, but they do not disappear, unless $|a|,|b|\rightarrow +\infty $.
In this latter case, the kernel $K(x,y)$ undergoes such rapid oscillations
in the $x$, or $y$ directions that it becomes negligible on the average and
cannot sustain bound states any more. In this limit, $|T|\rightarrow 1$ and $%
|R|\rightarrow 0$.

As far as bound states are concerned, the structure of Eq. (\ref{detM_1})
shows that only when $c_{V}=0$ $\det M_{+}$ does not depend on $E$, but on $%
\overline{k}$ only, so that, if $\overline{k}_{bs}$ is a solution of $\det
M_{+}\left( \overline{k}\right) =0$, both energies $\epsilon _{bs}=\pm \sqrt{%
m^{2}-\overline{k}_{bs}^{2}}$ are acceptable. This is shown in Fig.3, where $%
\det M_{+}\left( \overline{k}\right) =0$ is solved graphically for a scalar
well of strength $c_{S}=-m,$ $c=d=1$ and various values of $a=b$. With
increasing the latter phases, the two bound state energies quickly tend to
the thresholds of continuum, $\pm m$. For instance, when $a=b=10$, $\epsilon
_{bs}=\pm 0.999999923m$ and, in the continuum of scattering states, the
potential is almost reflectionless. 
\begin{figure}[h]
\begin{center}
\includegraphics[width=12cm,angle=0,clip]{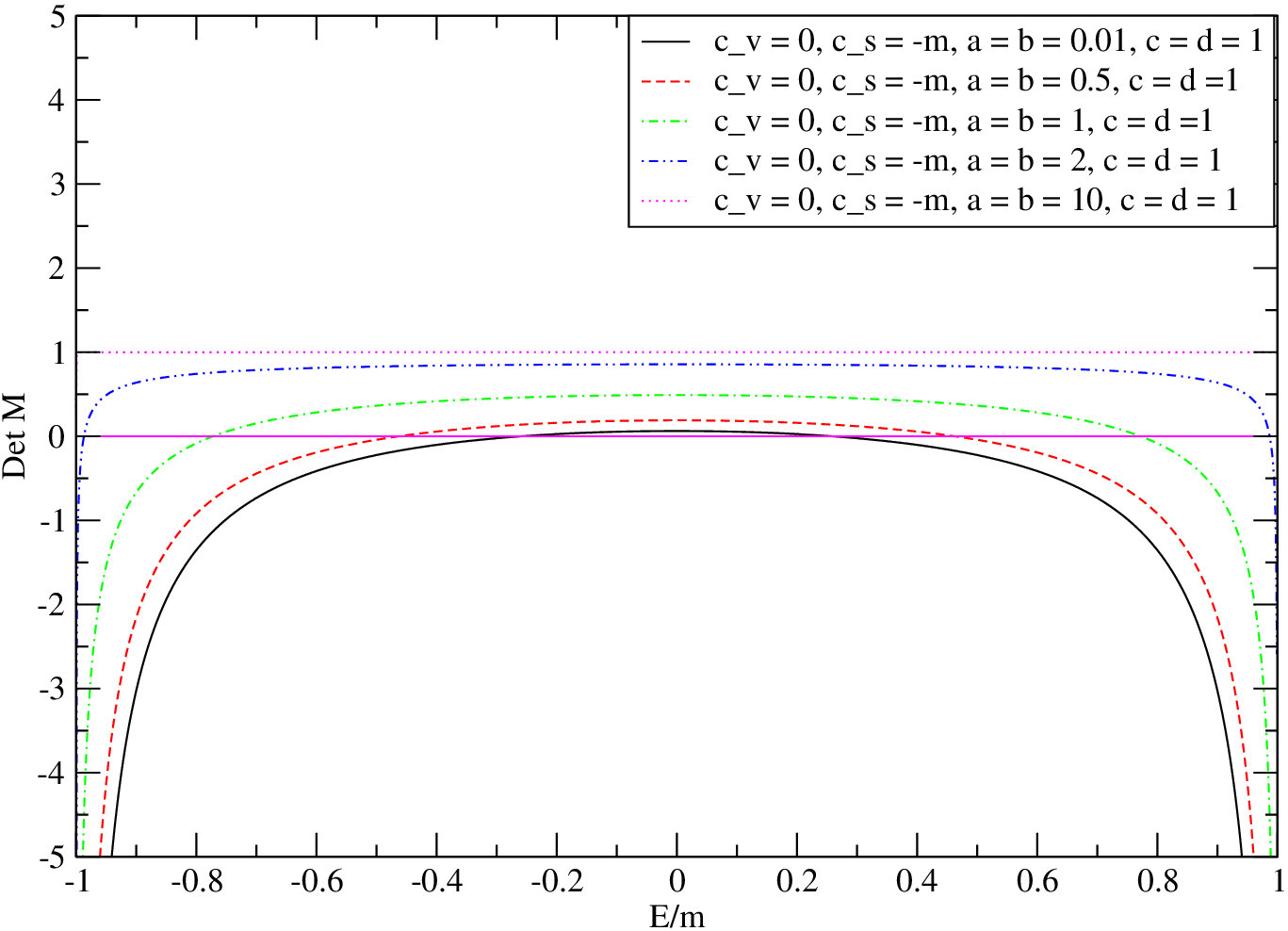}
\end{center}
\caption{Denominator of the transmission coefficient for a scalar well with $%
c_{S}=-m$, $c=d=1$ at various values of $a=b$.}
\label{Fig3}
\end{figure}

\section{Conclusions and perspectives}

In this work we have studied non-local $\mathcal{PT}$-symmetric potentials
in the one-dimensional Dirac equation. Owing to the fact that the definition
of the $S$ matrix adopted in our previous work\cite{CDV07} dedicated to
non-relativistic quantum mechanics is valid also in the relativistic case
(see e.g. Ref.\cite{W95}), we have used in the present work general
properties of the $S$ matrix under $\mathcal{P}$, $\mathcal{T}$ and $%
\mathcal{PT}$ transformations derived in Ref.\cite{CDV07}. There are, of
course, kinematical differences between Schr\"odinger and Dirac
formulations: in the latter case, total energies $E$ can be either positive
or negative; scattering states have either $E/m \leq -1$ or $E/m > +1$,
while bound states are found in the interval $-1<E/m<+1$.

The separable potential we have studied is very flexible, since, for
instance, it permits determining the real vector strength $c_{V}$ ( with
scalar strength $c_{S}=0$) that yields a bound state at an energy $E$
arbitrarily chosen in the $[-m,+m]$ interval (see Eqs.(\ref{detM_zero}-\ref%
{detM_1})).

Moreover, starting from the real kernel with real coupling strengths $c_V$
and $c_S$ and $a=b=0$, one can extend it in a natural way to the generalized
Hermitian case, with $g=h$ and $a=-b$, and, finally, to the $\mathcal{PT}$%
-symmetric case, with $g$ and $h$ even functions of their arguments and
arbitrary $a$ and $b$.

The specific choice of form factors $g(x)=e^{-c|x|}$ and $h(y)=e^{-d|y|}$
yields in the non-relativistic case transmission and reflection coefficients
that are rational functions of momentum, $k$, since they can be written as
ratios of polynomials in $k$. This opens the way to an algebraic search for
zeros of denominators, providing information on bound states, and of
numerators, e. g. of reflection coefficients ( transparency at given
momentum $k$), or transmission coefficients ( total reflectiveness at given $%
k$).

In the relativistic case the functional dependence is more involved, due to
the square root dependence on $k$ of energy $E=\sqrt {k^2 + m^2}$.
Nevertheless, it is interesting to remark that, in addition to the study of
properties of $T$ and $R$ at given $c_V$ and $c_S$, one can study specific
properties like absence of reflection or of transmission at given $k$ as
functions of $c_V$ and $c_S$: this can be easily done since transmission and
reflection coefficients are, respectively, second order polynomial in $c_V$
and/or $c_S$ over second order polynomial and first order over second order.

Study of the zeros of the denominators has already been mentioned in
connection with bound states. In the present work, we have made an effective
approach to $\mathcal{PT}$ symmetry, allowing for unitarity breaking of the
scattering matrix. The search for a Hermitian equivalent description would
imply the definition of a charge conjugation operator $\mathcal{C}$, in the
spirit of Ref.\cite{Be07}, or a metric operator $\eta _{+}$, according to
Ref.\cite{Mo08} and the study would be far from trivial. To our knowledge, $%
\eta _{+}$ in relativistic problems involving scattering states has been
exactly determined until now only for a non-Hermitian form of the
Klein-Gordon equation, either free\cite{Mo03}, or with a minimally coupled
electromagnetic field\cite{MZ06}.

This kind of more fundamental study, however, would be more appropriate to
finite-range potentials with exact $\mathcal{PT}$ symmetry, i.e. having a
purely real discrete spectrum with eigen-functions that are eigenstates of $%
\mathcal{PT}$ and reflectionless in the continuum, as discussed in ref.\cite%
{CDV07}. This could not be pursued for non-local potentials, but it could
work for \ the $\mathcal{PT}$-symmetric generalization of local scalar, or
pseudoscalar reflectionless potentials, like those constructed in Refs.\cite%
{TNZ93}-\cite{NT98}.

\appendix

\section{Appendix}

Here we explain the connection between the time-independent Green functions
used in the present work and the time dependent ones, which are solutions to
the equation%
\begin{equation}
\left( -i\alpha _{x}\frac{\partial }{\partial x}+\beta m-i\frac{\partial }{%
\partial t}\right) G\left( x,t;x^{\prime },t^{\prime }\right) =\delta \left(
x-x^{\prime }\right) \delta \left( t-t^{\prime }\right) \;.  \label{A1}
\end{equation}

We know from textbooks\cite{BS59} that particular solutions to Eq. (\ref{A1}%
) are the retarded component of the causal Green function%
\begin{equation}
G_{ret.}^{c}\left( x,t;x^{\prime },t^{\prime }\right) =\theta \left(
t-t^{\prime }\right) G_{+}\left( x-x^{\prime },t-t^{\prime }\right)
\label{A2}
\end{equation}%
and the advanced component%
\begin{equation}
G_{adv.}^{c}\left( x,t;x^{\prime },t^{\prime }\right) =\theta \left(
t^{\prime }-t\right) G_{-}\left( x-x^{\prime },t-t^{\prime }\right) \;,
\label{A3}
\end{equation}%
where $\theta \left( \tau \right) =1$ for $\tau >0$ and $\theta \left( \tau
\right) =0$ for $\tau <0$.

By inserting formulae (\ref{A2}-\ref{A3}) into Eq. (\ref{A1}), we obtain%
\begin{equation}
\begin{array}{c}
\theta \left( \pm \left( t-t^{\prime }\right) \right) \left( -i\alpha _{x}%
\frac{\partial }{\partial x}+\beta m-i\frac{\partial }{\partial t}\right)
G_{\pm }\left( x-x^{\prime },t-t^{\prime }\right) \\ 
+i\delta \left( t-t^{\prime }\right) G_{\pm }\left( x-x^{\prime
},t-t^{\prime }\right) =\delta \left( x-x^{\prime }\right) \delta \left(
t-t^{\prime }\right) \;\;,%
\end{array}
\label{A4}
\end{equation}%
where we have exploited the well-known relation $\frac{\partial }{\partial t}%
\theta \left( \pm \left( t-t^{\prime }\right) \right) =\delta \left(
t-t^{\prime }\right) $. Let us multiply both sides of Eq. (\ref{A4}) by $%
\exp \left( i\mathcal{E}\left( t-t^{\prime }\right) \right) $ and integrate
them over $u\equiv t-t^{\prime }$ from -$\infty $ to +$\infty $, with $%
\mathcal{E}$ a complex number whose imaginary part is chosen in such a way
that the integral exists: we must assume $\mathcal{E}_{+}\mathcal{=}%
E+i\epsilon $ for $G_{+}$ and $\mathcal{E}_{-}\mathcal{=}E-i\epsilon $ for $%
G_{-}$, with $\epsilon >0$. Thus, $G_{+}$ satisfies the equation%
\begin{equation}
\int_{0}^{+\infty }due^{i\mathcal{E}_{+}u}\left( -i\alpha _{x}\frac{\partial 
}{\partial x}+\beta m-i\frac{\partial }{\partial u}\right) G_{+}\left(
x-x^{\prime },u\right) -iG_{+}\left( x-x^{\prime },0\right) =\delta \left(
x-x^{\prime }\right) \;.  \label{A5}
\end{equation}

The above equation can be simplified by integrating by parts the third
integral on the left-hand side%
\begin{eqnarray*}
-i\int_{0}^{+\infty }due^{i\mathcal{E}_{+}u}\frac{\partial }{\partial u}%
G_{+}\left( x-x^{\prime },u\right) &=&-i\left\vert e^{i\mathcal{E}%
_{+}u}G_{+}\left( x-x^{\prime },u\right) \right\vert _{u=0}^{u=+\infty } \\
-\mathcal{E}_{+}\int_{0}^{+\infty }due^{i\mathcal{E}_{+}u}G_{+}\left(
x-x^{\prime },u\right) &=&iG_{+}\left( x-x^{\prime },0\right) -\mathcal{E}%
_{+}\int_{0}^{+\infty }due^{i\mathcal{E}_{+}u}G_{+}\left( x-x^{\prime
},u\right)
\end{eqnarray*}%
and the function%
\begin{equation}
G_{+}\left( x-x^{\prime }\right) \equiv \int_{0}^{+\infty }due^{i\mathcal{E}%
_{+}u}G_{+}\left( x-x^{\prime },u\right) =\int_{0}^{+\infty
}due^{iEu-\epsilon u}G_{+}\left( x-x^{\prime },u\right) \;,  \label{A6}
\end{equation}%
which is nothing but the Laplace transform of $G_{+}\left( x-x^{\prime
},t-t^{\prime }\right) $ with respect to time, satisfies the equation%
\begin{equation}
\left( -i\alpha _{x}\frac{\partial }{\partial x}+\beta m-\left( E+i\epsilon
\right) \right) G_{+}\left( x-x^{\prime }\right) =\delta \left( x-x^{\prime
}\right) \;  \label{A7}
\end{equation}%
and can be identified with the Green function corresponding to the complex
energy $E+i\epsilon $ for the time-independent Dirac equation.

We can proceed in the same way for $G_{-}\left( x-x^{\prime },t-t^{\prime
}\right) $, after introducing the complex energy $\mathcal{E}%
_{-}=E-i\epsilon $%
\begin{equation}
\int_{-\infty }^{0}due^{i\mathcal{E}_{-}u}\left( -i\alpha _{x}\frac{\partial 
}{\partial x}+\beta m-i\frac{\partial }{\partial u}\right) G_{-}\left(
x-x^{\prime },u\right) +iG_{-}\left( x-x^{\prime },0\right) =\delta \left(
x-x^{\prime }\right) \;.
\end{equation}

After integrating by parts the third integral on the l.h.s. of the above
equation and defining the Laplace transform with respect to time of $%
G_{-}\left( x-x^{\prime },t^{\prime }-t\right) $%
\begin{equation}
G_{-}\left( x-x^{\prime }\right) \equiv \int_{-\infty }^{0}due^{i\mathcal{E}%
_{-}u}G_{-}\left( x-x^{\prime },u\right) =\int_{0}^{+\infty
}dve^{-iEv-\epsilon v}G_{-}\left( x-x^{\prime },-v\right) \;,  \label{A8}
\end{equation}%
we arrive at the equation satisfied by the time-independent Green function $%
G_{-}\left( x-x^{\prime }\right) $%
\begin{equation}
\left( -i\alpha _{x}\frac{\partial }{\partial x}+\beta m-\left( E-i\epsilon
\right) \right) G_{-}\left( x-x^{\prime }\right) =\delta \left( x-x^{\prime
}\right) \;.  \label{A9}
\end{equation}

Eqs. (\ref{A7}-\ref{A9}) coincide with Eq. (\ref{time_ind_Green}) of the
text.

\section{Appendix }

Formula (\ref{detM_pm_rel}) can be easily proved from definition (\ref%
{detMpm}), according to which%
\begin{equation}
\begin{array}{c}
\det M_{+}-\det M_{-}=i\frac{c_{V}E+c_{S}m}{k}\left( S_{+}+S_{-}\right) +%
\frac{c_{V}^{2}-c_{S}^{2}}{4}\left(
D_{+}^{2}-D_{-}^{2}-S_{+}^{2}+S_{-}^{2}\right) \\ 
=i\frac{c_{V}E+c_{S}m}{k}\left( N_{+}^{\left( 1\right) }+N_{+}^{\left(
2\right) }+N_{-}^{\left( 1\right) }+N_{-}^{\left( 2\right) }\right) +\left(
c_{V}^{2}-c_{S}^{2}\right) \left( N_{-}^{\left( 1\right) }N_{-}^{\left(
2\right) }-N_{+}^{\left( 1\right) }N_{+}^{\left( 2\right) }\right) \;,%
\end{array}
\label{B1}
\end{equation}

where integrals $N_{\pm }^{\left( 1\right) }$ and $N_{\pm }^{\left( 2\right)
}$ are defined by formulae (\ref{N1pm}) and (\ref{N2pm}) respectively. We
promptly obtain from the definitions%
\begin{equation}
\begin{array}{c}
N_{+}^{\left( 1\right) }+N_{+}^{\left( 2\right) }+N_{-}^{\left( 1\right)
}+N_{-}^{\left( 2\right) }=2\int_{-\infty }^{+\infty }dxh\left( x\right)
e^{ibx}\int_{-\infty }^{+\infty }dx^{\prime }g\left( x^{\prime }\right)
e^{iax^{\prime }}\cos \left( k\left( x-x^{\prime }\right) \right) \theta
\left( x-x^{\prime }\right) \\ 
+2\int_{-\infty }^{+\infty }dxh\left( x\right) e^{ibx}\int_{-\infty
}^{+\infty }dx^{\prime }g\left( x^{\prime }\right) e^{iax^{\prime }}\cos
\left( k\left( x-x^{\prime }\right) \right) \theta \left( x^{\prime
}-x\right) \\ 
=\int_{-\infty }^{+\infty }dxh\left( x\right) e^{ibx}\int_{-\infty
}^{+\infty }dx^{\prime }g\left( x^{\prime }\right) e^{iax^{\prime }}\left(
e^{ik\left( x-x^{\prime }\right) }+e^{-ik\left( x-x^{\prime }\right) }\right)
\\ 
=\widetilde{g}\left( a-k\right) \widetilde{h}\left( b+k\right) +\widetilde{g}%
\left( a+k\right) \widetilde{h}\left( b-k\right) \;.%
\end{array}
\label{B2}
\end{equation}

and 
\begin{equation}
\begin{array}{c}
N_{-}^{\left( 1\right) }N_{-}^{\left( 2\right) }-N_{+}^{\left( 1\right)
}N_{+}^{\left( 2\right) }=-\widetilde{g}\left( a-k\right) \widetilde{h}%
\left( b+k\right) \widetilde{g}\left( a+k\right) \widetilde{h}\left(
b-k\right) \\ 
+\widetilde{g}\left( a-k\right) \widetilde{h}\left( b+k\right) N_{-}^{\left(
1\right) }+\widetilde{g}\left( a+k\right) \widetilde{h}\left( b-k\right)
N_{-}^{\left( 2\right) }\;.%
\end{array}
\label{B3}
\end{equation}

In deriving the last expression, the relation $\theta \left( -x\right)
=1-\theta \left( x\right) $ has been used in the integrands. Inserting the
right-hand sides of formulae (\ref{B2}-\ref{B3}) into formula (\ref{B1})
yields formula (\ref{detM_pm_rel}) of the text.

It is worthwhile to point out that the definitions we have used and,
consequently, the relation between $\det M_{+}$ and $\det M_{-}$ are also
valid for complex $k$, in particular for $k=i\overline{k,}$ with $\overline{k%
}>0$, characterizing bound states with real energy.

\end{document}